\documentclass[11pt,a4paper]{article}
\pdfoutput=1

\usepackage{graphicx}
\usepackage{appendix}
\usepackage{amsmath,amsfonts,amssymb,setspace}
\usepackage{braket,xcolor}
\usepackage{caption}
\usepackage{subcaption}
\usepackage{hyperref}
\usepackage{cite}
\usepackage{tocloft}
\usepackage{dsfont}
\usepackage[normalem]{ulem}
\usepackage[margin=0.75in]{geometry}
\hypersetup{colorlinks=false, linkcolor=blue, citecolor=red}

\def\be{\begin{eqnarray}}
	\def\ee{\end{eqnarray}}

\newcommand{\vs}[1]{\vspace{#1 mm}}
\begin{document}
	\begin{flushright}
		
		%UM-TH-00-16\\
		
		%SINP/TNP/00-20\\
		
		%arXiv:YYMM.NNNN\\
		
		%  
	\end{flushright}
	
	%\preprint{}
	
	\begin{center}
		{\Large{\bf Operator Complexity for Quantum Scalar Fields and Cosmological Perturbations}}\\
		\vs{10}
		%{\large
			{\large 
				S. Shajidul Haque${}^{a,\,}$\footnote{\url{shajid.haque@uct.ac.za}},\, Chandan Jana ${}^{b,\,}$\footnote{\url{channdann.jana@gmail.com}}, \,
				Bret Underwood${}^{c,\,}$\footnote{\url{bret.underwood@plu.edu}}}
			\vskip 0.3in
			{\it ${}^{a}$
				High Energy Physics, Cosmology \& Astrophysics Theory Group \\ and \\ The Laboratory for Quantum Gravity \& Strings \\ Department of Mathematics and Applied Mathematics, \\ University of Cape Town, South Africa }\vskip .5mm
			{\it ${}^{b}$ Mandelstam Institute for Theoretical Physics, Witwatersrand University, Johannesburg, South Africa}\vskip .5mm
			{\it ${}^{c}$ Department of Physics, Pacific Lutheran University, Tacoma, WA 98447}
			\vskip.5mm
		\end{center}
		\vskip.5mm
		%\end{center}
		\vskip 0.35in

\begin{abstract}
We calculate the operator complexity for the displacement, squeeze and rotation operators of a quantum harmonic oscillator. The complexity of the time-dependent displacement operator is constant, equal to the magnitude of the coherent state parameter, while the complexity of unitary evolution by a generic quadratic Hamiltonian is proportional to the amount of squeezing and is sensitive to the time-dependent phase of the unitary operator. We apply these results to study the complexity of a free massive scalar field, finding that the complexity has a period of rapid linear growth followed by a saturation determined by the UV cutoff and the number of spatial dimensions. We also study the complexity of the unitary evolution of quantum cosmological perturbations in de Sitter space, which can be written as time-dependent squeezing and rotation operators on individual Fourier mode pairs. The complexity of a single mode pair at late times grows linearly with the number of e-folds, while the complexity at early times oscillates rapidly due to the sensitivity of operator complexity to the phase of unitary time evolution. Integrating over all modes, the total complexity of cosmological perturbations scales as the square root of the (exponentially) growing volume of de Sitter space, suggesting that inflation leads to an explosive growth in complexity of the Universe.
\end{abstract}

%\begin{document}
%\maketitle
%\raggedbottom
%%~~~~~~~~~~~~~~~~~~~~~~~~~~~~~~~~~~~~~~~~~~~~~~~~~~~~~~~~~~~~~~~~~~~~~~~~~~~~~~~~~~~~~~~~~~
%%~~~~~~~~~~~~~~~~~~~~~~~~~~~~~~~~~~~~~~~~~~~~~~~~~~~~~~~~~~~~~~~~~~~~~~~~~~~~~~~~~~~~~~~~~~

\newpage

\tableofcontents

\section{Introduction}
\label{sec:intro}
%~~~~~~~~~~~~~~~~~~~~~~~~~~~~

As a concept, quantum circuit complexity is a measure of the number of simple quantum operations necessary to build a given unitary target operator -- which transforms some reference state into a target state -- from the identity.
For quantum systems in which the set of available operators form a continuum, quantum circuit complexity (hereafter denoted simply as complexity) can be interpreted in a geometric way as the length of a minimal geodesic in the space of operators \cite{NL1,NL2,NL3}.
Because of their wide-ranging applications, we are interested in the class of quantum systems 
built from the quantum harmonic oscillator.
In order to make progress in finding the complexity of a target operator,
it is often necessary to 
characterize the action of the operator on a restricted subspace of reference and target states, such as Gaussian wavefunctions,
as in e.g.~\cite{Jefferson,Guo:2018kzl,MyersMixed,me1,Ali:2019zcj}.
Such restrictions, however, make it difficult to make general statements about target unitaries, and can be difficult to extend to quantum field theory or interacting systems.

In this paper, we will consider the complexity of the target unitary operator directly, following the group manifold approach of \cite{Balasubramanian:2019wgd,Balasubramanian:2021mxo,Bai:2021ldj,Basteiro:2021ene} (see also \cite{Auzzi:2020idm}) and extending it to %infinite-dimensional Hilbert spaces.
the quantum harmonic oscillator.
In this formulation, the target unitary is generated from a set of fundamental operators, which form a Lie algebra.
Upon the choice of a suitable right-invariant metric, the minimal geodesic on the corresponding group manifold of the Lie algebra is found as a solution to the Euler-Arnold equation \cite{EulerArnold,Balasubramanian:2019wgd,Balasubramanian:2021mxo}.
The advantage of this group-theoretic approach is that the geometry is determined by the generators of the Lie algebra (up to the choice of the metric),
and is manifestly independent of the reference and target states.
Further, 
this approach can potentially be 
%readily 
extended to the study of complexity for interacting systems 
by suitably generalizing the Lie algebra of the fundamental operators beyond those considered here.

The target unitaries we consider here will be the displacement, squeeze and rotation operators of a quantum harmonic oscillator.
These operators are not only a starting point towards applying this formalism for complexity to more complex systems, they are also interesting from the perspective of continuous variable quantum information \cite{Braunstein:2005zz,Liu_2016,Zhuang:2019jyq}.
In particular, coherent and squeezed states of light are ubiquitous ingredients in quantum optics \cite{Schumaker}, and may play a supporting role in enhancing quantum computation algorithms (see e.g.~\cite{Braunstein:2005zz,Liu_2016,qumode}).
Squeezed states also show up in interesting applications in quantum field theories on curved backgrounds (such as \cite{Grishchuk,Albrecht,Martin1,Martin2}), and these states and the corresponding operators that generate them can lead to new quantum information theoretic perspectives on phenomena such as cosmological perturbations and Hawking radiation \cite{cosmology1, cosmology2,haque2020squeezed,Haque:2021kdm,Adhikari:2021ked}.
Since the complexity of these operators has been studied in state-based approaches before, we are able to compare our state-independent to computing complexity with those methods.

Complexity is particularly interesting when applied to many-body systems, where it may serve as a diagnostic for quantum chaos \cite{Maldacena2016-mb,Brown:2017jil,Camargo_2019,webme,Ali:2019zcj,Ryuchaos,Bhattacharyya:2020art,haqueReducedDensityMatrix} or topological phase transitions \cite{Ali:2018aon,PhysRevResearch.2.013323,Xiong:2019xoh}. To this end, the operator complexity of a single quantum harmonic oscillator is readily extended to a free massive quantum scalar field by writing the field as a sum over (decoupled) Fourier modes up to a UV cutoff. While still integrable, this simple system can serve as a testing ground for many-body operator complexity as well as uncover general features of the UV divergences that appear in complexity when moving to quantum field theory.
Another particularly interesting model is the unitary evolution of quantum cosmological perturbations in the presence of a time-dependent background.
At leading order, the Fourier modes of quantum cosmological perturbations behave as quantum harmonic oscillators with time-dependent frequencies \cite{Mukhanov}; the corresponding unitary time evolution operator can be written as a product of time-dependent squeezing and rotation operators \cite{Grishchuk,Albrecht,Martin1,Martin2,haque2020squeezed}.
The state complexity of cosmological perturbations has been studied before \cite{cosmology1, cosmology2,Adhikari:2021ked}, 
but being state-based, these calculations all require some specific choices for reference and target states.
Operator complexity instead is independent of the particular reference and target states, resulting in a more general statement about the complexity of the unitary evolution of the cosmological background.

The rest of the paper is organized as follows.
In Section \ref{sec:OperatorComplexity} we review the state-independent method of operator complexity from \cite{Balasubramanian:2019wgd,Balasubramanian:2021mxo}.
In Section \ref{sec:Heis} we apply this technique to the displacement operator,
finding that the operator-space geometry is 3-dimensional hyperbolic space, and the complexity is simply proportional to the magnitude of the coherent state parameter of the displacement operator.
In Section \ref{sec:OP_SU11} we consider a generic unitary that is quadratic in the creation and annihilation operators, which can be decomposed as a product of the squeeze and rotation operators.
After describing the geometry, we explore the upright and inverted harmonic oscillators as simple examples before finding the complexity for a generic product of squeeze and rotation operators.
In Section \ref{sec:Applications} we extend our analysis to quantum field theories with a UV cutoff. Specifically, we first consider a massive free scalar field, calculating the resulting complexity of unitary time evolution and its leading UV divergence.
We then analyze the complexity of unitary evolution of quantum cosmological perturbations in a de Sitter background for individual Fourier modes as well as the integral over all Fourier modes, comparing our results to corresponding calculations of state complexity.
In Section \ref{sec:Discussion}, we conclude with a discussion of our results and future directions.
Finally, in Appendix \ref{app:AltRep} we demonstrate that our results for the complexity for these groups is independent of the choice of matrix representation.

\section{Operator Complexity}
\label{sec:OperatorComplexity}
%%~~~~~~~~~~~~~~~~~~~~~~~~~~~~~~~~~~~

We start by reviewing the continuum circuit complexity construction we will use throughout the rest of the paper; see also \cite{Balasubramanian:2019wgd,Balasubramanian:2021mxo}.
Our quantum circuit is a unitary operator that transforms a given reference state $|\psi\rangle_{\rm R}$ to a specified target state $|\psi\rangle_{\rm T}$
\be
|\psi\rangle_{\rm T} = \hat {\mathcal U}_{\rm target}\ |\psi\rangle_{\rm R}
\label{eq:CircuitDef}
\ee
The target unitary $\hat {\mathcal U}_{\rm target}$ consists of a continuum of operations parameterized by a parameter $`s'$ that controls the level of the circuit
\be
\hat {\mathcal U}_{\rm target} = {\mathcal P}\, {\rm exp}\left[-i \int_0^1 V^I(s) \hat {\mathcal O}_I\ ds\right]\, ,
\label{eq:UtargetDef}
\ee
where the operators $\{\hat{\mathcal O}_I\}$ are some set of fundamental operators, the $V^I(s)$ are vectors that specify the path of the sequence of operators, 
and the path-ordering ${\mathcal P}$ ensures that the operators are applied sequentially from $s = 0$ to $s = 1$.
It is convenient to introduce the $s$-dependent unitary
\be
\hat U(s) = {\mathcal P}\, {\rm exp}\left[-i \int_0^s V^I(s') \hat {\mathcal O}_I\ ds'\right]\, ,
\label{eq:UsDef}
\ee
which is a solution to the differential equation
\be
\frac{d\hat U(s)}{ds} = -i V^I(s)\ \hat {\mathcal O}_I\ \hat U(s)\, ,
\label{eq:UsDiffeq}
\ee
subject to the boundary conditions 
\be
\hat U(0) = \mathds{1}\, \hspace{.2in} \mbox{and} \hspace{.2in} \hat U(1) = \hat {\mathcal U}_{\rm target}\, .
\label{eq:UsBoundary}
\ee

In principle, there are many different paths (``circuits'') $V^I(s)$ that can be used to build the target unitary through (\ref{eq:UsDiffeq}).
In order to identify the ``optimal'' path, we will characterize each path realizing the unitary (\ref{eq:UsDef}) by its circuit depth\footnote{There are other choices for the	cost function of the circuit depth \cite{NL1,NL2,NL3,Jefferson}. We will choose the geodesic cost function for its relative simplicity and geometric interpretation.}
\be
{\mathcal D}\left[V^I\right] = \int_0^1 \sqrt{G_{IJ} V^I V^J}\ ds\, ,
\label{eq:CircuitDepth}
\ee
corresponding to the geodesic length in the space of operators.
The metric $G_{IJ}$ identifies the operational ``cost'' or weight to building the path with any particular operator $\hat {\mathcal O}_I$; as we will discuss, while a natural choice is the Cartan-Killing form of the Lie algebra of the operators $\{\hat {\mathcal O}_I\}$, $G_{IJ} = K_{IJ}$, leading to a bi-invariant metric, this choice will not be possible for the non-compact groups we are interested in here, as we will discuss in the following sections.
Instead, for the examples we consider in this paper we will choose a flat metric on the operators $G_{IJ} = \delta_{IJ}$, leading to a right-invariant metric, 
so that there are no preferred directions in operator space (though we leave $G_{IJ}$ as arbitrary for the remaining parts of this section, for clarity).
More generally, the metric may be constructed either phenomenologically, by including the difficulty of preparing a particular gate in the lab, or by including a theoretical bias for ``simple'' versus ``composite'' operators.
The choice of metric $G_{IJ}$, together with (\ref{eq:UsDiffeq}), leads to a notion of distance on this space \cite{Jefferson}
\be
ds^2 = G_{IJ}\ dV^I dV^J
\label{eq:OperatorMetric}
\ee
where
\be
dV^I = {\rm Tr}\left( d\hat U\ \hat U^{-1}\ \hat {\mathcal O}_I^\dagger\right)\, .
\label{eq:OperatorLocalCoord}
\ee

The optimal quantum circuit is the one with minimal circuit depth, so that the complexity of (\ref{eq:UtargetDef}) is the minimization of the circuit depth
\be
{\mathcal C}_{\rm target} = \min_{\{V^I\}} {\mathcal D}\left[V^I\right] = \min_{\{V^I\}}\int_0^1 \sqrt{G_{IJ} V^I V^J}\ ds\, ,
\label{eq:ComplexityDef}
\ee
over all possible paths $\{V^I(s)\}$ realizing the target operator (\ref{eq:UtargetDef}).
The minimal path $V^I(s)$ is therefore a geodesic on the space (\ref{eq:OperatorMetric}), which solves the Euler-Arnold equation
\cite{Balasubramanian:2019wgd,Balasubramanian:2021mxo,EulerArnold}
\be
G_{IJ} \frac{dV^J}{ds} = f_{IJ}^P\ V^J G_{PL} V^L\, ,
\label{eq:EulerArnold}
\ee
where the $f_{IJ}^P$ are the structure constants of the operators,
\be
\left[\hat {\mathcal O}_I, \hat {\mathcal O}_J\right] = i f_{IJ}^P\ \hat {\mathcal O}_P\, .
\ee

In the following sections, we will proceed as follows.
First, we identify the target unitary $\hat {\mathcal U}_{\rm target}$ and select a set of basis operators $\{\hat {\mathcal O}_I\}$, with associated Lie group, that we use to construct this unitary. 
Paths that solve (\ref{eq:EulerArnold}) define a set of geodesics $\{V^I(s)\}$ on this space.
We then restrict this set of geodesics to those that realize the target unitary through (\ref{eq:UsDiffeq}) and the boundary conditions (\ref{eq:UsBoundary}).
Finally, we use the resulting optimal construction of the unitary to calculate the complexity (\ref{eq:ComplexityDef}), and determine its dependence on the parameters of the target unitary.

\section{Displacement Operator Complexity}
\label{sec:Heis}

As a simple application, we begin by analyzing the circuit complexity associated with the \emph{displacement operator}
\be
\hat {\mathcal U}_{\rm target} = \hat {\mathcal D}(\alpha) = {\rm exp} \left[\alpha \hat a^\dagger - \alpha^* \hat a\right]\, ,
\label{eq:DispTargetU1}
\ee
which generates the coherent state $|\alpha\rangle = \hat {\mathcal D}(\alpha) |0\rangle$ from the vacuum.
More generally, we can consider the \emph{time-dependent displacement operator}
\be
\hat {\mathcal U}_{\rm target}(t) = \hat {\mathcal D}(\alpha,t) = e^{i\hat H_0 t} \hat {\mathcal D}(\alpha) e^{-i\hat H_0 t} = \hat {\mathcal D}\left(\alpha e^{-i\omega t}\right)\, ,
\label{eq:DispTargetU2}
\ee
in which the displacement operator (\ref{eq:DispTargetU1}) is time-evolved by a free Hamiltonian $\hat H_0 = \omega \hat a^\dagger \hat a$.
The displacement operator is a standard starting point for many approaches to calculating complexity \cite{Guo:2018kzl,Guo:2020dsi,Bhattacharyya:2020art,Caputa:2021sib}, and as such will provide a good place to start in applying the formalism developed in Section \ref{sec:OperatorComplexity} as applied to infinite-dimensional Hilbert spaces.

A natural set of fundamental operators to use for generating (\ref{eq:DispTargetU1}) through the construction (\ref{eq:UtargetDef}) 
\be
\hat {\mathcal D}(\alpha, t) = {\mathcal P}\ {\rm exp} \left[-i \int_0^1 V^I(s) \hat e_I\ ds\right]\, ,
\label{eq:DispTargetU3}
\ee
are the Hermitian operators
\be
\hat e_1 = \frac{1}{\sqrt{2}} \left(\hat a + \hat a^\dagger\right), \hspace{.2in} \hat e_2 = \frac{i}{\sqrt{2}} \left(\hat a - \hat a^\dagger\right), \hspace{.2in} \hat e_3 = \hat {\mathds 1}\, ,
\ee
which obey the standard Heisenberg Lie algebra $[\hat e_1,\hat e_2] = -i \hat e_3$, with all other commutators vanishing.
As a result, the only non-zero structure constant (up to permutations) is $f_{12}^3 = -1$, so that the Cartan-Killing form vanishes $K_{IJ} = f_{IK}^L f_{JL}^K = 0$.
In terms of the $\{\hat e_I\}$, we can write our target operator as
\be
\hat {\mathcal U}_{\rm target}(t) = {\rm exp} \left[\sqrt{2} i\ {\rm Im}[\alpha(t)]\ \hat e_1 + \sqrt{2} i\ {\rm Re}[\alpha(t)]\ \hat e_2\right]\, .
\label{eq:DispTargetU4}
\ee

The $s$-dependent vectors $V^I(s)$ in (\ref{eq:DispTargetU3}) parameterize the path through the space of operators, from the identity (at $s=0$) to the target operator (at $s=1$).
The associated circuit depth (\ref{eq:CircuitDepth}) measures the geodesic length of any particular path, given a metric $G_{IJ}$ on the space of operators.
Since the Cartan-Killing form vanishes, a natural choice for the metric is a diagonal metric $G_{IJ} = \delta_{IJ}$ so that the infinitesimal circuit depth $G_{IJ} V^I V^J = \left(V^1\right)^2 + \left(V^2\right)^2 + \left(V^3\right)^2$ is non-zero.

The path from the identity to our target state with the smallest circuit depth is thus obtained as a solution to the Euler-Arnold equation (\ref{eq:EulerArnold}), which becomes
\be
\frac{dV^1}{ds} &=& - V^2 V^3\, ; \nonumber \\
\frac{dV^2}{ds} &=& V^1 V^3\, ; \label{eq:HeisenbergEulerArnold}\\
\frac{dV^3}{ds} &=& 0\, . \nonumber 
\ee
Combining the first two equations of (\ref{eq:HeisenbergEulerArnold}), it is straightforward to see that $(V^1)^2 + (V^2)^2$ is independent of $s$.
Since the third equation of (\ref{eq:HeisenbergEulerArnold}) leads to $V^3(s) = c_3$, a constant, we have the following general solution for a depth-minimizing path
\be
V^1(s) &=& v_1 \cos(v_3 s) + v_2 \sin(v_3 s)\, ; \nonumber \\
V^2(s) &=& v_1 \sin(v_3 s) - v_2 \cos(v_3 s) \, ; \label{eq:HeisEASolution}\\
V^3(s) &=& v_3\, ; \nonumber 
\ee
where $v_1,v_2, v_3$ are constants that will be determined by boundary conditions.
The resulting circuit complexity along this minimal path is then simply
\be
{\mathcal C}_{\rm target} = \int_0^1 \sqrt{G_{IJ} V^I(s) V^J(s)}\ ds = \sqrt{v_1^2+v_2^2+v_3^2}\, .
\ee

Before imposing the boundary conditions to fix the constants $\{v_I\}$ in terms of the target unitary operator, let us make a simplification in our construction of the unitary operator in terms of the $\hat e_I$.
Since the operator $\hat e_3$ is a center that commutes with all other elements, its contribution to (\ref{eq:DispTargetU3}) is just an overall time-independent phase when operating on states
\be
{\rm exp} \left[ -i \int_0^1 V^I \hat e_I\ ds \right]\, |\psi\rangle_{\rm R} &=& {\rm exp} \left[-i \int_0^1 \left(V^1(s) \hat e_1 + V^2(s) \hat e_2\right) ds\right] e^{-i v_3 \hat {\mathds 1}}\, |\psi\rangle_{\rm R} \nonumber \\
&=& {\rm exp} \left[-i \int_0^1 \left(V^1(s) \hat e_1 + V^2(s) \hat e_2\right) ds\right]\, e^{-i v_3}\, |\psi\rangle_{\rm R}\, .
\ee
We will therefore set $v_3 = 0$ to avoid this additional phase ambiguity.
The resulting solutions (\ref{eq:HeisEASolution}) for the depth-minimizing path 
simplify so that the solutions (\ref{eq:HeisEASolution})
become $V^1(s) = v_1, V^2(s) = v_2, V^3(s) = 0$ with 
corresponding complexity
\be
{\mathcal C}_{\rm target} = \sqrt{v_1^2 + v_2^2}\, .
\label{eq:HeisIntermediateComplexity}
\ee

Our target operator (\ref{eq:DispTargetU1}) is the $s=1$ boundary condition of the $s$-dependent unitary operator (\ref{eq:UsDef})
\be
\hat U(s) = {\mathcal P}\, {\rm exp}\left[-i \int_0^s V^I(s') \hat e_I\ ds'\right]\, ,
\label{eq:UsDefHeis}
\ee
which is a solution to the differential equation (\ref{eq:UsDiffeq})
\be
\frac{d\hat U(s)}{ds} = -i V^I(s)\ \hat e_I\ \hat U(s)\, .
\label{eq:UsDiffeqHeis}
\ee
In order to find an explicit solution to (\ref{eq:UsDiffeqHeis}), we will use an 
explicit 3 x 3 upper-triangular matrix representation\footnote{We obtain identical results for other matrix representations, indicating that the precise choice of matrix representation is not important. See Appendix \ref{app:AltRep}.} of the Heisenberg Lie algebra generators as
\be
\hat e_1 = \begin{pmatrix} 
		0 & 1 & 0 \cr
		0 & 0 & 0 \cr
		0 & 0 & 0
	\cr\end{pmatrix}\, , \hspace{.2in}
\hat e_2 = \begin{pmatrix} 
	0 & 0 & 0 \cr
	0 & 0 & 1 \cr
	0 & 0 & 0
	\cr\end{pmatrix}\, , \hspace{.2in}
\hat e_3 = \begin{pmatrix} 
	0 & 0 & i \cr
	0 & 0 & 0 \cr
	0 & 0 & 0
	\cr\end{pmatrix}\, .
\ee
Within this representation, a general element of the Heisenberg group can be written in terms of group elements $a,b,c$
\be
\hat U(s) = \begin{pmatrix}
		1 & i a(s) & c(s) \cr
		0 & 1 & i b(s) \cr
		0 & 0 & 1
	\end{pmatrix}\, .\label{eq:HeisGeneralU}
\ee
Following (\ref{eq:OperatorLocalCoord}), we can use (\ref{eq:HeisGeneralU}) and the explicit representation of the 
fundamental operators $\{\hat e_i\}$ to construct the operator-space geometry of the Heisenberg group
\be
ds^2 = G_{IJ}\ dV^I\ dV^J = da^2 + db^2 + \left(b da + dc\right)^2\, ,
\ee
which has constant negative curvature.
The operator-space geometry associated with the Heisenberg group is thus $3$-dimensional hyperbolic space.

With the explicit representation (\ref{eq:HeisGeneralU}) and the solutions for the $V^I(s)$, 
the differential equation (\ref{eq:UsDiffeqHeis}) is solved by the parameterizations
\be
a(s) &=& a_0 - v_1 s\, ; \nonumber \\
b(s) &=& b_0 - v_2 s\, ; \\
c(s) &=& c_0 + b_0 v_1 s - \frac{1}{2} v_1 v_2 s^2\, , \nonumber
\ee
where the $a_0, b_0, c_0$ are constants.
Imposing the boundary condition that the unitary reduce to the identity operator at $s=0$, $\hat U(s=0) = \hat {\mathds 1}$ leads to the unitary
\be
\hat U(s) = \begin{pmatrix}
		1 & -iv_1 s & -\frac{1}{2} v_1 v_2 s^2 \cr
		0 & 1 & -i v_2 s \cr
		0 & 0 & 1
	\end{pmatrix}\, .
\ee
The boundary condition at $s=1$, $\hat U(s=1) = \hat {\mathcal U}_{\rm target}$, with (\ref{eq:DispTargetU4}), now leads to a solution for the constants
$v_1 = -\sqrt{2}\ {\rm Im}[\alpha(t)], v_2 = -\sqrt{2}\ \rm{Re}[\alpha(t)]$.
The resulting complexity for the time-dependent displacement operator is thus 
\be
{\mathcal C}_{\rm Heis} = \sqrt{2}\, |\alpha|\, ,
\label{eq:DisplacementComplexity}
\ee
and is simply proportional to the time-independent magnitude of the coherent state parameter.
This result for the complexity of the displacement operator is similar to previous results \cite{Guo:2018kzl,Guo:2020dsi} for the complexity of a corresponding coherent state, and stands in contrast to other measures of complexity of the displacement operator \cite{Bhattacharyya:2020art,Caputa:2021sib}, in which the complexity is time-dependent.

It is interesting to interpret the result (\ref{eq:DisplacementComplexity}) in terms of the average
number density -- or equivalently, the average energy -- of a vacuum coherent state
\be
\langle E \rangle \sim \bar{N}_\alpha = \langle \alpha | \hat a^\dagger \hat a | \alpha\rangle = |\alpha|^2 \, .
\label{eq:CoherentStateE}
\ee
From this perspective, the complexity of the displacement operator (\ref{eq:DisplacementComplexity}) scales as the square root of average energy (average number of particles) 
\be
{\mathcal C}_{\rm displacement} \sim \sqrt{\langle E\rangle}\, .
\label{eq:DispComplexEnergy}
\ee
In realizations of quantum information protocols, the energy needed to prepare a state or set of gates can be an important resource \cite{Braunstein:2005zz,qumode}.
Because of the scaling (\ref{eq:DispComplexEnergy}), 
%the complexity of a coherent state grows only as the square root of the energy needed to construct the state.
%Alternatively, the relation (\ref{eq:DispComplexEnergy}) 
this implies that the energy required to build a coherent state with some fixed complexity ${\mathcal C}_*$ grows quadratically with that complexity $\langle E \rangle \sim {\mathcal C}_*^2$.
It would be interesting to study further whether these scalings have general lessons for building quantum information protocols with continuous variables in the lab.

\section{Squeezing Operator Complexity}
%\section{SU(1,1) Operator Complexity}
\label{sec:OP_SU11}
%%~~~~~~~~~~~~~~~~~~~~~~~~~~~~~~~~~~~~~~~~~~~~

Let us now turn our attention to a unitary quantum circuit that is constructed from a generic quadratic combination of creation and annihilation operators
\be
|\psi\rangle_{\rm T} = \hat U_2|0\rangle\, ,
\label{eq:SMG}
\ee
where $\hat U_2$ is the unitary operator
\be
\hat U_2 = {\rm exp}\left[-i \int \hat H_2\ dt\right]\,
\label{eq:U2Def}
\ee
defined in terms of a quadratic Hamiltonian
\be
\hat H_2 = \Omega\ \hat a^\dagger \hat a + \frac{1}{2}\left(\Delta\, \hat a^2 + \Delta^*\ \hat a^{\dagger 2}\right)\, .
\label{eq:QuadHamiltonian}
\ee
The unitary operator (\ref{eq:U2Def}) can also be written more generally in the factorized form \cite{Schumaker}
\be
\hat U_2 = \hat S(r,\phi) \hat R(\theta)
\label{eq:U2Def2}
\ee
in terms of the squeeze and rotation operators
\be
\hat S(r, \phi) = e^{\frac{r}{2} (e^{- 2 i \phi} \hat a^2 - e^{2 i \phi} \hat a^{\dagger 2})}\hspace{.2in} \text{and}  \hspace{.2in}
\hat R(\theta) = e^{- i \theta  \frac{\hat a^{\dagger} \hat a+\hat a \hat a^\dagger}{2}}\, ,
\label{eq:SqueezeDef}
\ee
where $r$, the squeezing parameter, characterizes the amount of squeezing, $\phi$ is the squeezing angle, and $\theta$ is the rotation angle.
The formulation (\ref{eq:U2Def2}) in terms of the squeezing and rotation operators can be quite useful, particularly when the parameters $\Omega, \Delta$ of the target quadratic Hamiltonian (\ref{eq:QuadHamiltonian}) are time-dependent, as often happens for interesting physical applications.
Thus, we can represent our quantum circuit more generally as the \emph{squeezed state}
\be
|\psi\rangle_{\rm T} = \hat {\mathcal U}_{\rm target} |\psi\rangle_{\rm R} = \left[\hat S(r,\phi)\ \hat R(\theta)\right] |\psi\rangle_{\rm R}\, ,
\label{eq:QuantumCircuit}
\ee
where $|\psi\rangle_{\rm R}$ is an arbitrary reference state.
From another perspective, a Bogoluibov transformation between two sets $(\hat a, \hat a^\dagger)\leftrightarrow (\hat b, \hat b^\dagger)$ of creation and annihilation operators can be written as a transformation with respect to the squeeze and rotation operators
\be
\hat a = \alpha\, \hat b + \beta^*\, \hat b^\dagger = \hat {\mathcal U}_{\rm target}^\dagger\, \hat b\, \hat {\mathcal U}_{\rm target}\, .
\ee

From the form of $\hat H_2$, the natural set of fundamental operators $\{\hat {\mathcal O}_I\}$ to use in building the target unitary (\ref{eq:UtargetDef}) are the Hermitian operators
\be
\hat e_1 = \frac{\hat a^2+\hat a^{\dagger 2}}{4} \,, \qquad \hat e_2 = \frac{i(\hat a^2-\hat a^{\dagger 2})}{4} \,, \qquad \hat e_3 = \frac{\hat a\hat a^\dagger + \hat a^\dagger \hat a}{4} \, .
\label{eq:su11generators}
\ee
These operators satisfy the $\mathfrak{su}(1,1)$ Lie algebra,
\be 
\label{su11algebra}
[\hat e_1, \hat e_2] = - i \hat e_3,\ [\hat e_3, \hat e_1] &=&  i \hat e_2, \ 
[\hat e_2, \hat e_3] =  i\hat e_1 \,,
\ee
therefore the target operator $\hat {\mathcal U}_{\rm target}$ is a generic element of SU(1,1).
The structure constants $[\hat e_i, \hat e_j] = if_{ij}^k \hat e_k$
\be
f^3_{12}= -1, \  f^2_{31}= 1, f^1_{23}= 1
\label{eq:su11Fij}
\ee
define the $\mathfrak{su}(1,1)$ Cartan-Killing form $K_{ij}= f_{i k}^{l} f^k_{jl}$
\be
K= \begin{pmatrix}
	2 & \ \ 0 & \ 0\\ 
	0 &\  \ 2 & \ 0\\
	0 & \ \ 0 &  -2
\end{pmatrix}\,.  
\label{eq:su11Killing}
\ee

Following the discussion in Section \ref{sec:OperatorComplexity}, a natural choice for a metric on the space of operators of $\mathfrak{su}(1,1)$ would be proportional to the bi-invariant Cartan-Killing form $G_{ij} = K_{ij}$. However, because of the negative eigenvalue in (\ref{eq:su11Killing}) such a metric would be Lorentzian, potentially leading to a zero or imaginary complexity (\ref{eq:ComplexityDef}) for ``lightlike'' ($G_{ij}V^i V^j = 0$) or ``timelike'' ($G_{ij} V^i V^j < 0$) paths, respectively.
Since we would like to interpret the complexity (\ref{eq:ComplexityDef}) as a continuum version of the number of gates needed to build the target quantum circuit, negative or imaginary complexities are undesirable.
Instead, here we will again choose a diagonal Riemannian metric with equal cost factors $G_{ij} = \delta_{ij}$ so that $G_{ij} V^i V^j = \left(V^1\right)^2 + \left(V^2\right)^2 + \left(V^3\right)^2$ is non-negative.
This means that all of our fundamental operators are ``easy,'' in contrast to the SU(2) analysis of \cite{Balasubramanian:2019wgd} which chose one of the directions to be a ``hard'' direction with an increased cost factor.

Given our characterization of the target circuit in terms of the $\mathfrak{su}(1,1)$ generators $\{\hat e_i\}$
\be
\hat {\mathcal U}_{\rm target} = \hat S(r,\phi) \hat R(\theta) = {\mathcal P}\, {\rm exp}\left[-i \int_0^1 V^i(s)\ \hat e_i\ ds\right]\, ,
\label{eq:SqueezedTarget}
\ee
we now need to find paths $\{V^1(s), V^2(s), V^3(s)\}$ that minimize the circuit depth (\ref{eq:CircuitDepth}).
The geodesic equation (\ref{eq:EulerArnold}) becomes the set of three equations for the $V^i(s)$
\be
\frac{dV^1}{ds} &=& -2 V^2 V^3\, ;  \nonumber \\
\frac{dV^2}{ds} &=& 2 V^1 V^3\, ;  \label{eq:su11EulerArnold}\\
\frac{dV^3}{ds} &=& 0\, .\nonumber 
\ee
Solutions to (\ref{eq:su11EulerArnold}) take the form
\be
V^1(s) &=& v_1 \cos (2v_3 s) - v_2 \sin(2v_3 s) \,; \nonumber \\
V^2(s) &=& v_1 \sin (2v_3 s) + v_2 \cos(2v_3 s) \,; \label{eq:su11EASoln} \\
V^3(s) &=& v_3 \,,\nonumber 
\ee
where $v_1, v_2, v_3$ are constants that we will determine by matching the form of the target operator (\ref{eq:SqueezedTarget}).
With this solution (\ref{eq:su11EASoln}) the magnitude of the vector $V^i$ takes the simple form
\be
G_{ij} V^i V^j = \left(V^1\right)^2 + \left(V^2\right)^2 + \left(V^3\right)^2 = v_1^2 + v_2^2 +v_3^2 = v^2 + v_3^2
\ee
where we defined $v^2 = v_1^2 + v_2^2$ for convenience. The complexity (\ref{eq:ComplexityDef}), the minimal circuit depth evaluated on the solution to the Euler-Arnold equation (\ref{eq:su11EASoln}), becomes
\be
{\mathcal C}_{\rm target} = \sqrt{v^2 + v_3^2}\, .
\label{eq:su11Complexity1}
\ee

The constants $v_i$ are determined by matching to the target operator (\ref{eq:UsDef}), which solves (\ref{eq:UsDiffeq}) 
\be
\frac{d\hat U(s)}{ds} = -i V^I(s)\ \hat e_i\ \hat U(s)\, ,
\label{eq:su11Diffeq}
\ee
subject to the boundary conditions (\ref{eq:UsBoundary}) $\hat U(0) = \hat{\mathds{1}}$, $\hat U(1) = \hat {\mathcal U}_{\rm target}=\hat S(r,\phi) \hat R(\theta)$.
In order to solve (\ref{eq:su11Diffeq}), we work with the following 2 x 2 representation of the $\mathfrak{su}(1,1)$ generators,
%\footnote{Note that since SU(1,1) is infinite-dimensional, there is no finite-dimensional \emph{unitary} representation of this group, therefore we will choose this non-unitary 2 x 2 dimensional representation.},
\be 
\hat e_1= \frac{1}{2}\begin{pmatrix}
	0 &  -1 \\ 
	1 & \ 0
\end{pmatrix}, \ \hat e_2= \frac{1}{2}\begin{pmatrix}
	0 & i \\ 
	i & \ 0
\end{pmatrix}, \hat e_3= \frac{1}{2}\begin{pmatrix}
	1 & \ 0\\ 
	0 & -1
\end{pmatrix}\,.
\label{eq:su11representation}
\ee
%As we show in Appendix \ref{app:AltRep}, a 3 x 3 representation gives identical results.
Within the 2 x 2 representation, a general element of SU(1,1) takes the form
\be
\hat U(s) = \begin{pmatrix}
	q(s) & p(s)^\ast \\ p(s) & q(s)^\ast 
\end{pmatrix} 
\qquad \text{such that} \qquad |q|^2 -|p|^2 = 1 \,.
\label{eq:su11UAnsatz}
\ee
We can further write this in terms of some $\alpha, \beta, \gamma$, which parametrize the group
\be
\hat U(s) = \begin{pmatrix}
	\cosh(\rho) + i \gamma \frac{\sin(\rho)}{\rho} & (\alpha - i \beta) \frac{\sin(\rho)}{\rho} \\
	(\alpha + i \beta) \frac{\sin(\rho)}{\rho} & \cosh - i \gamma \frac{\sin(\rho)}{\rho}
\end{pmatrix}\, ,
\label{eq:su11GeneralU}
\ee
where $\rho^2 = \alpha^2 + \beta^2 - \gamma^2$.
Following (\ref{eq:OperatorLocalCoord}), we can use (\ref{eq:su11GeneralU}) and the explicit representation of the 
fundamental operators $\{\hat e_i\}$ to construct the non-compact operator-space geometry of the $\mathfrak{su(1,1)}$ generators
\be
ds^2_{\mathfrak{su(1,1)}} =&& 4 \frac{\sinh^4\rho}{\rho^4} \left[(\beta^2+\gamma^2 + \rho^2 \coth^2\rho)d\alpha^2+(\alpha^2+\gamma^2+\rho^2\coth^2\rho)d\beta^2  \right. \nonumber \\
&& \left. +(\alpha^2+\beta^2+ \rho^2 \coth^2\rho)d\gamma^2-2 \alpha \beta d\alpha d\beta - 2 (2 \beta \rho \coth \rho + \alpha \gamma)d\alpha d\gamma 
\right. \nonumber \\
&& \left. + 2 (2 \alpha \rho \coth\rho - \beta \gamma)d\beta d\gamma \right]\, . \label{eq:su11Metric}
\ee
The corresponding curvature
\be
R_{\mathfrak{su(1,1)}} = -8 (\rho^2 + 2\gamma^2) \tanh^2 \rho - \rho^2
\ee
is non-positive.

Returning to the form (\ref{eq:su11UAnsatz}), the differential equation (\ref{eq:su11Diffeq}) then becomes equivalent to the differential equations for $q(s)$ and $p(s)$
\be
\frac{dq}{ds} &=&\, -\frac{iv_3}{2} q + \frac{(v_2 + iv_1)}{2} e^{-2iv_3s} p\,; \\
\frac{dp}{ds} &=&\, \frac{iv_3}{2} p + \frac{(v_2 - iv_1)}{2} e^{2iv_3s} q\,,
\ee
with solutions
\be
q(s) &=& e^{-iv_3 s}(c_1 e^{\lambda s/2} + c_2 e^{-\lambda s/2}) \,; \\
p(s) &=& \frac{v_2-iv_1}{v^2} e^{iv_3 s} \left(c_1(\lambda-i v_3) e^{\lambda s/2} - c_2(\lambda+v_3) e^{-\lambda s/2}\right)\, ,
\ee
where again we used $v_1^2 + v_2^2 = v^2$ for notational simplicity, and $\lambda = \sqrt{v^2-v_3^2}$ can be either real or imaginary, depending on the relative sizes of $v$ and $v_3$.
Demanding the boundary condition $\hat U(0) = \hat{\mathds{1}}$ implies $q(0) = 1, p(0) = 0$, which fixes $c_1 =\frac{1}{2\lambda}(\lambda+i v_3), c_2 = \frac{1}{2\lambda}(\lambda-i v_3)$, so that $\hat U(s)$ becomes
\be
\hat U(s) &=&
\begin{pmatrix}
	e^{- i v_3s} \left(\cosh\left(\frac{\lambda s}{2}\right)+i\frac{v_3}{\lambda}\sinh\left(\frac{\lambda s}{2}\right)\right)  & \,\, e^{- i v_3s}\, \frac{(v_2+iv_1)}{\lambda} \sinh \left(\frac{\lambda s}{2}\right)   \\
	e^{i v_3s}\,\frac{(v_2-iv_1)}{\lambda} \sinh \left(\frac{\lambda s}{2}\right)  & \,\, 	e^{ i v_3s} \left(\cosh\left(\frac{\lambda s}{2}\right)-i\frac{v_3}{\lambda}\sinh\left(\frac{\lambda s}{2}\right)\right)
\end{pmatrix} \label{eq:su11UsSolution}
%&=& \begin{pmatrix}
%	e^{- i v_3s}  & 0 \\
%	0  & e^{ i v_3s}
%\end{pmatrix} 
%\begin{pmatrix}
%	\cosh (vs)  & \frac{v_2+iv_1}{v} \sinh (vs)  \\
%	\frac{v_2-iv_1}{v} \sinh (vs) & \cosh (vs)
%\end{pmatrix} \,,
\ee
%where we noted that $\hat U(s)$ can be factorized as above.
Imposing the remaining boundary condition at $s=1$, $\hat U(1) = {\mathcal U}_{\rm target}$ will allow us to determine the geodesic constants $v_i$ in terms of the target operator quantities.
However, before we do this in the general case, let us examine limiting cases in order to gain an intuition for the results.

\subsection{Free Harmonic Oscillator}
\label{sec:FreeHarmonic}

\begin{figure}[t]
\centering\includegraphics[width=.4\textwidth]{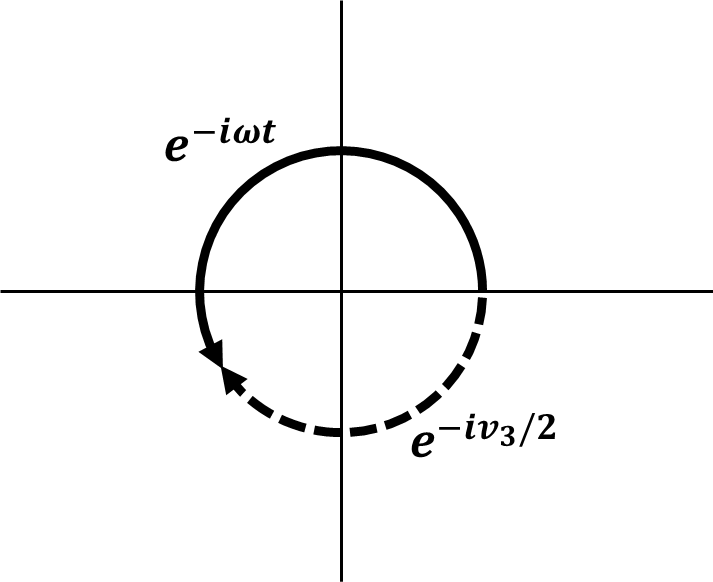}\includegraphics[width=.6\textwidth]{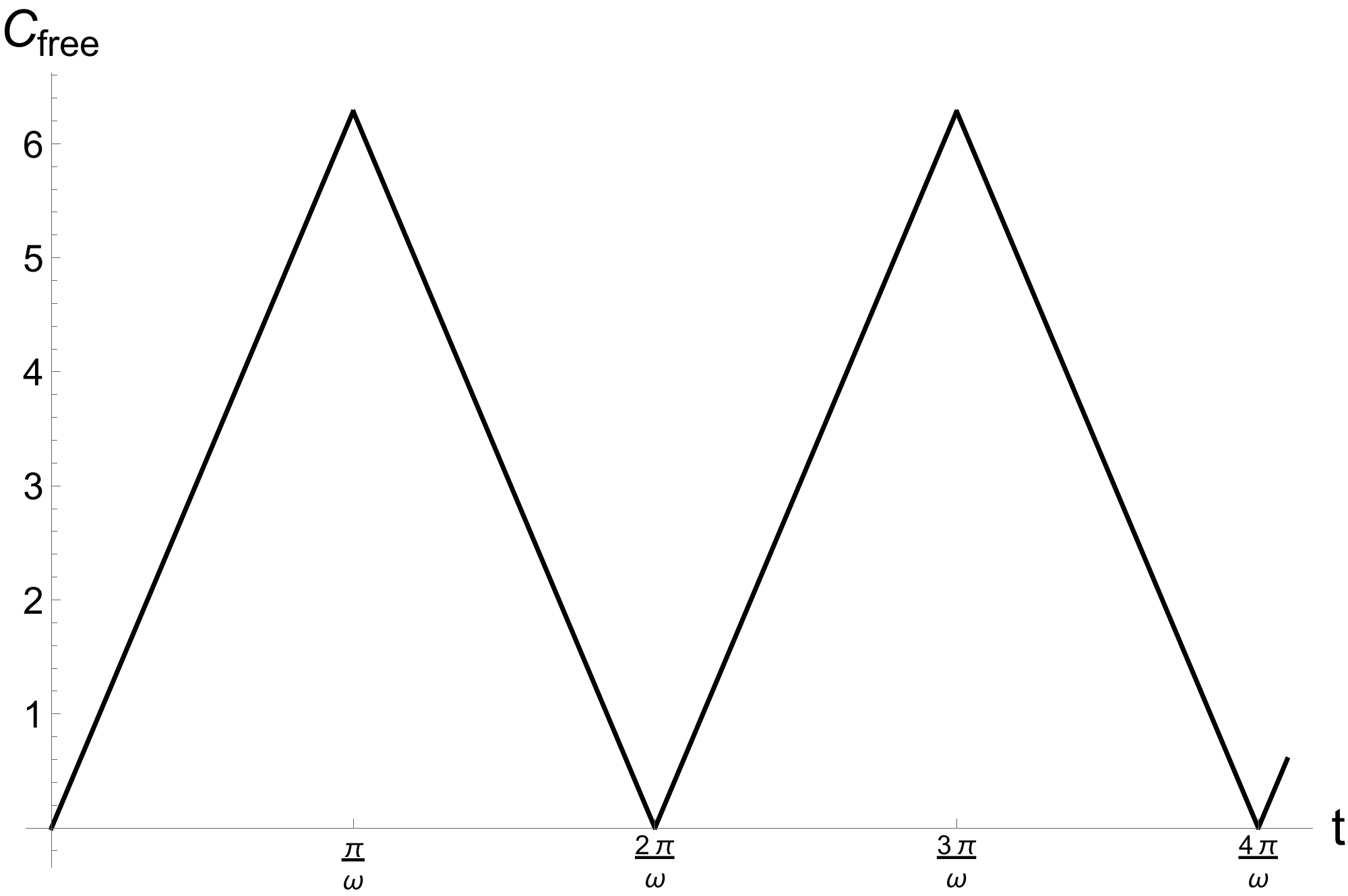}
\caption{(Left) When $\omega t > \pi$, the minimal length geodesic for $|v_3|$ to satisfy the boundary condition (\ref{eq:FreeV3}) becomes $|v_3| = 2(2\pi - \omega t)$. (Right) Because of this ``shortcut'', the complexity of the time evolution operator $e^{-i\hat H_0 t}$ for the free harmonic oscillator oscillates between positive and negative slopes $\pm 2\omega$ with a period given by the quantum revival time $T = 2\pi/\omega$.}
\label{fig:FreeComplexity}
\end{figure}

As our first example, let us consider time evolution under the free harmonic oscillator Hamiltonian as our target operator
\be
|\psi\rangle_{\rm T} = e^{-i\hat H_0 t} |\psi\rangle_{\rm R}
\ee
with $\hat H_0 = \frac{\omega}{2} (\hat a^\dagger \hat a + \hat a \hat a^{\dagger}) = 2\omega \hat e_3$.
In terms of the squeeze and rotation operators (\ref{eq:SqueezeDef}) this corresponds to vanishing squeezing $r = 0$ and a time-dependent rotation angle $\theta(t) = \omega t$.
The resulting target operator is then
\be
\hat {\mathcal U}_{\rm target} = e^{-i\hat H_0 t} = \begin{pmatrix}
		e^{-i\omega t} & 0 \\
		0 & e^{i\omega t}
		\end{pmatrix}\, .
	\label{eq:FreeUTarget}
\ee
Imposing the boundary condition $\hat U(1) = \hat {\mathcal U}_{\rm target}$ on the operator (\ref{eq:su11UsSolution}), we find
$v_1 = v_2 = 0$ so that $\lambda = i v_3$, and
\be
e^{-i\omega t} = e^{-i v_3/2}\, .
\label{eq:FreeV3}
\ee
Since the complexity becomes ${\mathcal C}_{\rm free} = \sqrt{v_1^2 + v_2^2 + v_3^2} = |v_3|$, the simple solution $v_3 = 2\theta(t) = 2\omega t$ to (\ref{eq:FreeV3}) results in a complexity that is unbounded and grows linearly with time ${\mathcal C}_{\rm free} = 2\omega t$.
However, the complexity should be a \emph{minimization} over the $v_i$, and this is not the minimal solution to (\ref{eq:FreeV3}).
For $\omega t > \pi$, we can solve (\ref{eq:FreeV3}) with the smaller angle $|v_3| = 2(2\pi-\omega t)$, 
up until $\omega t > 2\pi$, at which point $v_3 = 4(\omega t - 2\pi)/3$ 
(and so on for increasing windings).
The resulting complexity as a function of time oscillates between positive and negative slopes of $\pm 4\omega/3$, as seen in Figure \ref{fig:FreeComplexity}
Indeed, since the operator (\ref{eq:FreeUTarget}) is identical after the revival time $\hat {\mathcal U}_{\rm target}(t=0) = \hat {\mathcal U}_{\rm target}(t=T)$ for $T = 2\pi/\omega$, we expect the complexity to show a similar periodicity.

\subsection{Inverted Harmonic Oscillator}

Next we consider the inverted harmonic oscillator, with Hamiltonian
\be
\hat H_I = -\frac{\Omega}{2} \left(\hat a^2 + {\hat a}^{\dagger 2}\right) = -2\Omega\, \hat e_1\, .
\ee
Taking the time evolution operator as our target operator again, with representation
\be
\hat {\mathcal U}_{\rm target} = e^{-i\hat H_I t} = e^{2i \Omega t \hat e_1} = \begin{pmatrix}
	\cosh (\Omega t) & -i\sinh(\Omega t) \\
	i\sinh(\Omega t) & \cosh(\Omega t) 
\end{pmatrix}\, ,
\label{eq:InvertedTarget}
\ee
we see this as a squeeze operator with linearly increasing squeezing $r(t) = \Omega t$, constant squeeze angle $\phi = \pi/4$, and vanishing rotation angle $\theta = 0$.
Matching $\hat U(1) = \hat {\mathcal U}_{\rm target}$ leads to the conditions
\be
e^{-iv_3} \left(\cosh\left(\frac{\lambda}{2}\right)+i\frac{v_3}{\lambda}\sinh\left(\frac{\lambda}{2}\right)\right) \,  = \cosh(\Omega t) \,, \qquad \frac{v_2+iv_1}{\lambda}\sinh \left(\frac{\lambda}{2}\right) \, e^{-iv_3} = -i\sinh(\Omega t) \,,
\ee
which are solved by $\lambda = v_1 = 2 r(t) = 2 \Omega t$, $v_2 = v_3 = 0$. 
The resulting complexity for the inverted harmonic oscillator grows linearly with time
\be
{\mathcal C}_{\rm Invert} = 2 \Omega t\, .
\label{eq:InvertedComplexity}
\ee
Unlike the free harmonic oscillator of the previous subsection, the complexity for the inverted harmonic oscillator grows without bound, reflecting the instability of the inverted oscillator.

\subsection{General Squeezing and Rotation}
\label{sec:SqueezeRotComplexity}

\begin{figure}[t]
	\centering
	\begin{subfigure}[b]{0.49\textwidth}
		\centering \includegraphics[width=\textwidth]{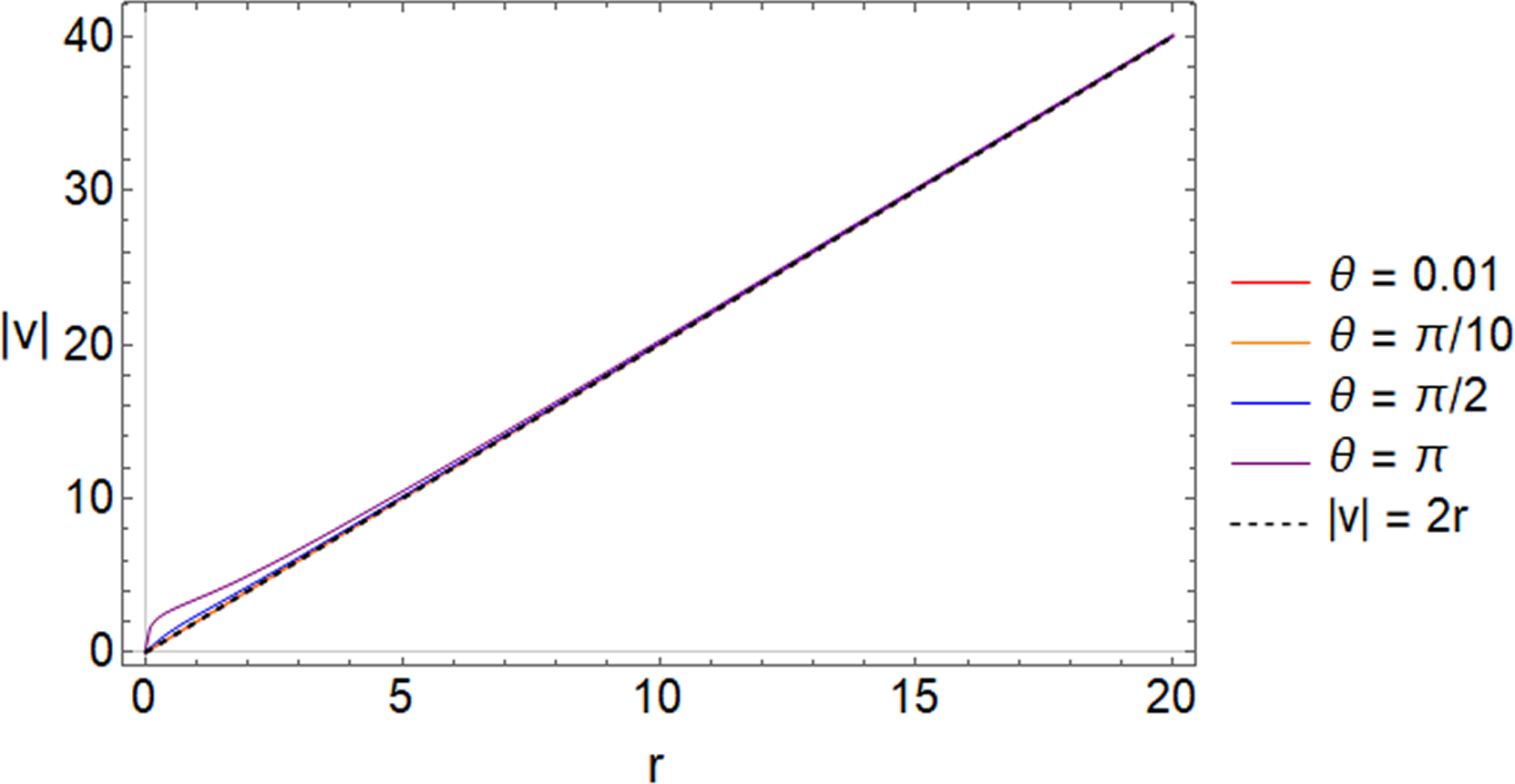}
		\caption{}
		\label{subfig:vNumerics}
	\end{subfigure} \hfill
	\begin{subfigure}[b]{0.49\textwidth}
		\centering \includegraphics[width=\textwidth]{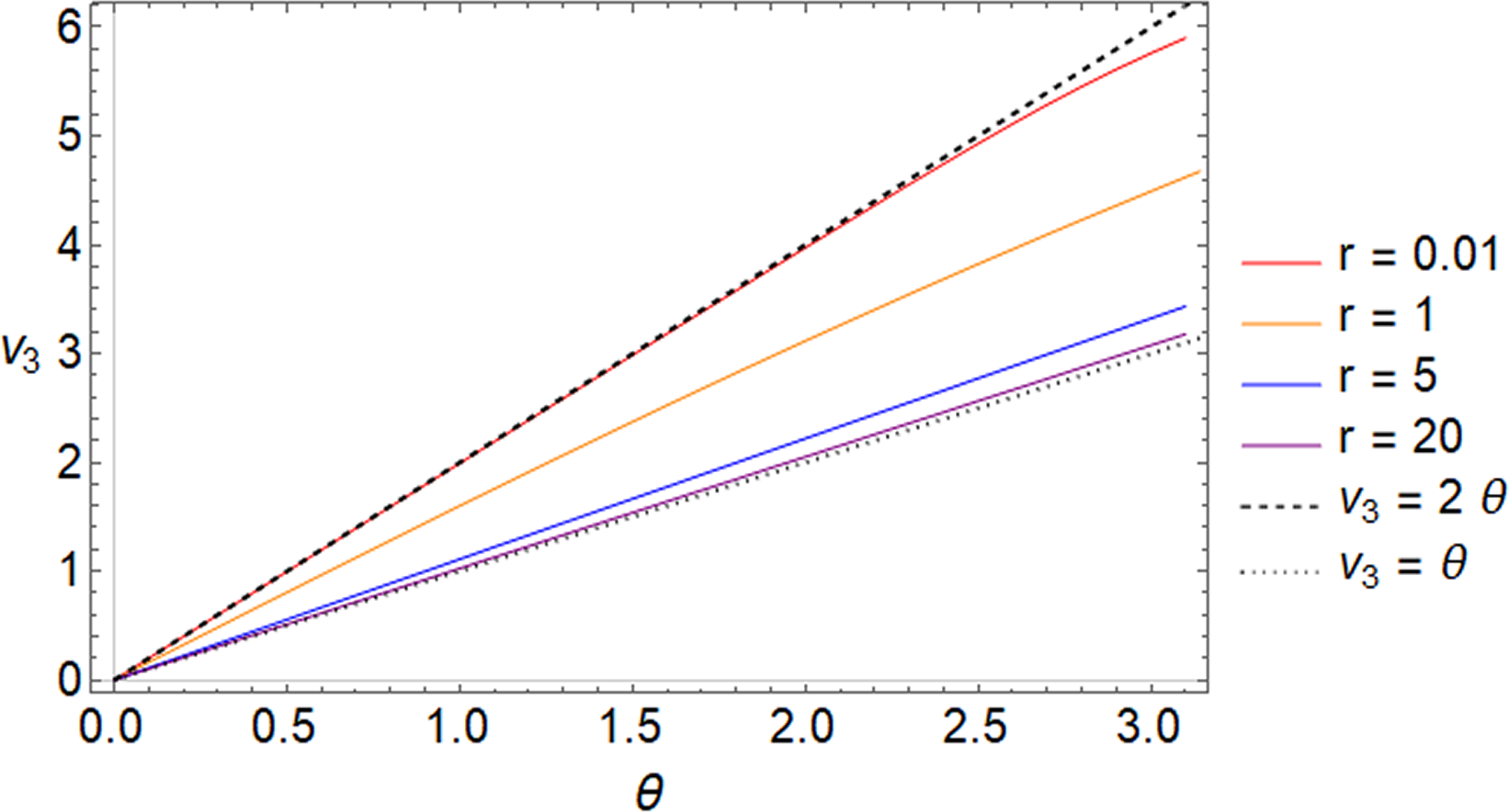}
		\caption{}
		\label{subfig:v3Numerics}
	\end{subfigure}
	\caption{Numerical solutions to the boundary condition (\ref{eq:SqueezeMatch}) for the geodesic parameters (a) $v = \sqrt{v_1^2 + v_2^2}$, illustrating that $v \approx 2r$, and (b) $v_3$, illustrating the behavior (\ref{eq:SU11BoundarySolutionApprox}).}
	\label{fig:SU11Numerics}
\end{figure}

Having spent some time on the simplified special cases of the previous two subsections, we now consider a more general element of SU(1,1).
The most general target operator can be written as a product of squeeze and rotation operators
\be
\hat {\mathcal U}_{\rm target} = \hat S(r,\phi) \hat R(\theta)\, .
\label{eq:Squeeze}
\ee
As discussed above, a generic quadratic time-evolution operator (\ref{eq:U2Def}) can always be decomposed into a product of this form, in which the parameters $r(t), \phi(t), \theta(t)$ all inherit time-dependence through the Heisenberg equation of motion.
Using the forms of the squeezing and rotation operators (\ref{eq:SqueezeDef}), and the generators (\ref{eq:su11generators}) with the representation (\ref{eq:su11representation}), we write our target operator as
\be
\hat {\mathcal U}_{\rm target} = \begin{pmatrix}
	e^{-i\theta} \cosh r & e^{i(2\phi+\theta)} \sinh r \\ e^{-i(2\phi+\theta)} \sinh r & e^{i\theta} \cosh r 
\end{pmatrix}\, .
\label{eq:SqueezeTarget}
\ee
Matching the operator (\ref{eq:su11UsSolution}) at the boundary condition $\hat U(1) = \hat {\mathcal U}_{\rm target}$ with (\ref{eq:SqueezeTarget}) we obtain the conditions
\be
%\cosh v \, e^{-iv_3} = e^{-2i\theta} \cosh r \,, \qquad  \frac{v_2+iv_1}{v}\sinh v \, e^{-iv_3} = e^{2i(\theta+\phi)} \sinh r \,,
e^{-i\theta}\cosh r = e^{-iv_3} \left(\cosh\left(\frac{\lambda}{2}\right)+i\frac{v_3}{\lambda}\right)\, , \qquad e^{-i(2\phi+\theta)}\sinh r =  \frac{e^{iv_3}(v_2-i v_1)}{\lambda}\sinh\left(\frac{\lambda}{2}\right)\, .
\label{eq:SqueezeMatch}
\ee
%with solutions
%\be
%v &=& r\, ; \\
%|v_3| &=& \begin{cases}
%	\theta - 2\pi n & \mbox{ for } 2\pi n < \theta < 2\pi (n+1) \cr
%	2\pi n - \theta & \mbox{ for } 2\pi (n-1) < \theta < 2\pi n
%	\end{cases}\mbox{ for some } n\, ,
%\label{eq:SqueezeSolution}
%\ee
%where the solution for $v_3$ reflects the oscillation of the minimization procedure discussed in Section \ref{sec:FreeHarmonic}.
Solutions to (\ref{eq:SqueezeMatch}) for $v = \sqrt{v_1^2 + v_2^2} = v(r,\theta), v_3 = v_3(r,\theta)$ as functions of the squeezing parameter and rotation angle can be found analytically in the limits of small $r \ll 1$ and large $r \gg 1$ squeezing\footnote{The relative sizes of $v_1$ and $v_2$ are determined from the squeezing angle $\phi$ from (\ref{eq:SqueezeMatch}), but will not be important in the rest of our analysis.}:
\be
v &\approx & 2 r \, ; \\
|v_3| &=& \begin{cases} 2 \theta_{\rm min} & \mbox{ for } r \ll 1 \cr
	\theta_{\rm min} & \mbox{ for } r \gg 1
	\end{cases}
\label{eq:SU11BoundarySolutionApprox}
\ee
where 
\be
\theta_{\rm min} = \begin{cases}
	\theta -  2\pi n & \mbox{ for } 2\pi n < \theta < \pi (2n + 1) \cr
	2\pi n - \theta & \mbox{ for } \pi (2n-1) < \theta < 2\pi n
	\end{cases} \mbox{ for some } n \, ,
\ee
represents the minimization of the rotation angle discussed in Section \ref{sec:FreeHarmonic}.
More generally, the equations (\ref{eq:SqueezeMatch}) can be solved numerically as seen in Figure \ref{fig:SU11Numerics}, illustrating that this analytic solution holds quite generally, with $v_3$ interpolating between $2\theta_{\rm min}$ and $\theta_{\rm min}$ for intermediate values of the squeezing $r$.

The resulting complexity of a generic squeezed state is thus given by
\be
{\mathcal C}_{\rm Squeeze} \approx \sqrt{4r^2 + v_3^2}\, ,
\label{eq:CSqueeze1}
\ee
where $v_3$ is given by (\ref{eq:SU11BoundarySolutionApprox}).
Since $v_3$ is bounded by $2\pi$, for large squeezing $r \gg 1$ the complexity is linear in the squeezing parameter
\be
{\mathcal C}_{\rm Squeeze} \approx 2r\, .
\label{eq:CSqueeze2}
\ee
Interestingly, the complexity of the target operator (\ref{eq:Squeeze}) consisting of the product of the squeeze and rotation operators depends periodically on the rotation angle $\theta$ through $v_3$, but is entirely independent of the squeezing angle $\phi$.

The complexity of a \emph{Gaussian} squeezed state, generated by acting the squeezing operator (\ref{eq:Squeeze}) on a vacuum state, has been considered before using different techniques.
For example, in \cite{cosmology1} the complexity of a single-mode squeezed state was calculated by parameterizing the effect of a squeezing operator on Gaussian wavefunctions, using the approach of \cite{Jefferson,me1}.
In contrast to our results above (\ref{eq:CSqueeze2}) and (\ref{eq:CSqueeze2}), the complexity of a single-mode squeezed state in \cite{cosmology1} depends sensitively on the squeezing angle. Further, the Gaussian wavefunction approach of \cite{Jefferson, me1} ignores normalization factors and phases of the Gaussian wavefunction, and cannot be sensitive to the rotation angle $\theta$ in the way found in (\ref{eq:CSqueeze1}), even though such states are technically distinct.
The differences between the two approaches is heightened by applying them to the inverted harmonic oscillator.
Our result above (\ref{eq:InvertedComplexity}) finds that the complexity is equal to the time-dependent squeezing parameter $r(t) = 2\Omega t$, growing linearly with time, while the complexity for the Gaussian wavefunction approach \emph{saturates} at late times ${\mathcal C}_{\rm Gauss} \approx \pi/4$ \cite{cosmology1}.
These differences may simply be due to the fact that the pure operator complexity, as we have found here, is independent of the specific forms of the reference and target states, and is thus the complexity of the quantum circuit as it would be applied to an \emph{arbitrary} reference state.
A more specific reference state, such as a vacuum state as considered in \cite{cosmology1}, perhaps allows one to find additional shortcuts by using the Gaussian form of the position-space wavefunction.
In this way, our results for the operator complexity (\ref{eq:CSqueeze1}) may serve as a universal upper bound on the complexity needed to construct the target operator as applied to an arbitrary reference state. 
Specific choices of reference and target states can then allow one to find shortcuts in the construction of the operator with smaller circuit depth.

The complexity of a quantum circuit can also be calculated by characterizing the circuit through a covariance matrix \cite{Hackl:2018ptj}. Applying this approach to a squeezed state leads to a complexity proportional to the squeezing ${\mathcal C}_{\rm cov} = r(t)$ \cite{cosmology1,Lehners_2021}, similar to our results above (although the results \cite{cosmology1,Lehners_2021} appear again to be insensitive to the rotation angle $\theta$).
However, both the covariance matrix and Gaussian wavefunction approaches suffer from similar limitations: they can only be applied to Gaussian reference and target states. In contrast, the operator complexity approach we have outlined here is independent of the reference and target states, and is potentially generalizable to other groups that are not quadratic in the raising and lowering operators.

%%%%%%%%%%%%%%%%%%%%%%%%%%%%%%%%%%%%%%%%%%%%%%%%%%%%%%%%%%%%%
%%%%%%%%%%%%%%%%%%%%%%%%%%%%%%%%%%%%%%%%%%%%%%%%%%%%%%%%%%%%%
%%%%%%%%%%%%%%%%%%%%%%%%%%%%%%%%%%%%%%%%%%%%%%%%%%%%%%%%%%%%%
\section{Scalar Field Complexity}
\label{sec:Applications}

In the previous sections, we studied the operator complexity of the displacement operator as well as the free and and squeezed quantum harmonic oscillators.
Now we extend the formalism and techniques of operator complexity from \cite{Balasubramanian:2019wgd,Balasubramanian:2021mxo} to a quantum scalar field, first in the case of a free massive scalar field, then to quantum scalar cosmological perturbations.

\subsection{Free Scalar Field}
\label{subsec:FreeField}

To begin, let us consider a $(d+1)$-dimensional free scalar field of mass $m$ in a box of size $L$ with periodic boundary conditions (we will take $L\rightarrow \infty$ at the end).
Expanding the field in Fourier modes
\be
\hat \phi(x) = \sum_{\vec{n}}^\infty \frac{1}{\sqrt{2 E_{\vec{n}}}} \left(\hat a_{\vec{n}}\, e^{i\vec{p}_{\vec{n}}\cdot \vec{x}} + \hat a_{\vec{n}}^\dagger\, e^{-i \vec{p}_{\vec{n}}\cdot \vec{x}}\right)
\label{eq:scalarFourier}
\ee
for canonical creation and annihilation operators $\left[\hat a_{\vec{n}}, \hat a_{\vec{m}}^\dagger\right] = \delta_{\vec{n},\vec{m}}$,
where $\vec{p}_{\vec{n}} = \vec{n} \pi/L$ and $E_{\vec{n}} = \sqrt{\vec{p}_{\vec{n}}^2 + m^2}$ and $\vec{n}$ is a $d$-dimensional vector of integers.
The Hamiltonian becomes a sum over modes
\be
\hat H = \frac{1}{2}\sum_{n_1=1}^\infty \sum_{n_2 = 1}^\infty ... \sum_{n_d=1}^\infty E_{\vec{n}} \left(\hat a_{\vec{n}}^\dagger \hat a_{\vec{n}} + \hat a_{\vec{n}} \hat a_{\vec{n}}^\dagger\right) \, .
%= \sum_{\vec{n}}^\infty E_{\vec{n}} \hat e_3^{\vec{n}}
\ee
We will cutoff the infinite sums of modes at the UV scale $\Lambda = N_{\rm max} \pi/L$ for $N_{\rm max} \gg 1$, allowing us to study how the complexity diverges as a function of the UV cutoff $\Lambda$.
As a result, our target unitary 
%$\hat {\mathcal U}_{\rm target} = {\rm exp}\left[-i \hat H t\right]$ 
becomes the product
\be
\hat {\mathcal U}_{\rm target} = e^{-i\hat H t} = \prod_{n_1 = 1}^{N_{\rm max}} \prod_{n_2 = 1}^{N_{\rm max}}...\prod_{n_d = 1}^{N_{\rm max}} e^{-i \frac{1}{2}E_{\vec{n}} \left(\hat a_{\vec{n}}^\dagger \hat a_{\vec{n}} + \hat a_{\vec{n}} \hat a_{\vec{n}}^\dagger\right)}\, .
\label{eq:FreeTargetU}
\ee
For each mode $\vec{n}$ we can choose our set of fundamental gates to be the corresponding $\mathfrak{su(1,1)}$ generators from Section \ref{sec:OP_SU11}, e.g. $\hat e_3^{\vec{n}} = (\hat c_{\vec{n}}^\dagger \hat c_{\vec{n}} + \hat c_{\vec{n}} \hat c_{\vec{n}}^\dagger)/4$.
The corresponding geometry on the space of operators becomes the $(N_{\rm max})^d$ dimensional direct product of copies of the $\mathfrak{su(1,1)}$ geometry (\ref{eq:su11Metric})
\be
ds^2 = \sum_{n_1=1}^{N_{\rm max}} \sum_{n_2 = 1}^{N_{\rm max}}... \sum_{n_d=1}^{N_{\rm max}} G_{IJ}^{\vec{n}} V^I_{\vec{n}} V^J_{\vec{n}}\, .
\ee
For any single mode $\vec{n}$ of the target unitary (\ref{eq:FreeTargetU}), the minimal path, and corresponding complexity, is given by that of the free harmonic oscillator of Section \ref{sec:FreeHarmonic}, 
\be
{\mathcal C}_{\rm free}^{\vec{n}} = |v_3^{\vec{n}}| = \begin{cases}
	2(E_{\vec n}\, t - 2\pi m) & \mbox{for } 2\pi m < E_{\vec{n}}\, t< \pi (2m+1) \cr
	2(2\pi m - E_{\vec{n}}\,t) & \mbox{for } \pi (2m-1) < E_{\vec{n}}\, t < 2\pi m
	\end{cases} \mbox{ for some integer } m
\label{eq:modeComplexity}
\ee
which oscillates between $0 \leq {\mathcal C}_{\rm free}^{\vec{n}}  \leq 2\pi$. 
Taking into account all of the modes, the total complexity for the (regularized) free scalar field is
\be
{\mathcal C}_{\phi} = \sqrt{\sum_{\{n_i\}=1}^{N_{\rm max}} \left(v_3^{\vec{n}}\right)^2}\, ,
\label{eq:scalarfieldComplexity}
\ee
where $v_3^{\vec{n}}$ is given by (\ref{eq:modeComplexity})
Taking the continuum limit $L\rightarrow \infty$, the sums become integrals $\sum_{\{n_i\}=1}^{N_{\rm max}} \rightarrow L^d/\pi^d \int^\Lambda d^dp$, and the complexity (\ref{eq:scalarfieldComplexity}) becomes
\be
{\mathcal C_{\phi}} = \frac{L^{d/2}}{\pi^{d/2}} \sqrt{\int^\Lambda v_3(p)^2\, d^d p}\, 
\label{eq:scalarFieldComplexityPhi}
\ee
where $v_3(p)$ is the continuum version of (\ref{eq:modeComplexity}) with $E_p = \sqrt{\vec{p}^2 + m^2}$.
Even before performing the integral, we see that the scalar field complexity (\ref{eq:scalarFieldComplexityPhi}) diverges as the square root of the volume, similar to other field theory complexity calculations using a geometric cost function \cite{Jefferson}.

At very early times $\pi/t \gg \Lambda$ much shorter than the UV scale, none of the modes in (\ref{eq:scalarFieldComplexityPhi})
has yet reached its first oscillation in $v_3^{\vec{n}}$. Thus, we have $v_3(p) = 2E_p\, t$ in these early times, and the complexity becomes
\be
{\mathcal C}_{\phi} \sim \frac{L^{d/2}}{\pi^{d/2}} \sqrt{\int^\Lambda 4(p^2 + m^2)\, t^2\, d^dp}
 \sim \frac{L^{d/2}}{\pi^{d/2}} \left(\mbox{vol}(d)\right)^{1/2} \Lambda^{(d+2)/2} t \sim L^{d/2} \Lambda^{d/2} \Lambda t
\label{eq:phiEarlyComplexity}
\ee
where we took the integral to be dominated by the high-energy modes $p^2 \gg m^2$, and $\mbox{vol}(d)$ is the volume of a $d$-dimensional unit sphere.
The UV divergence at these early times scales as $\Lambda^{(d+2)/2}$, leading to a rapid growth of complexity over a very short time scale.

\begin{figure}[t]
	\centering
	\begin{subfigure}[b]{0.49\textwidth}
		\centering \includegraphics[width=\textwidth]{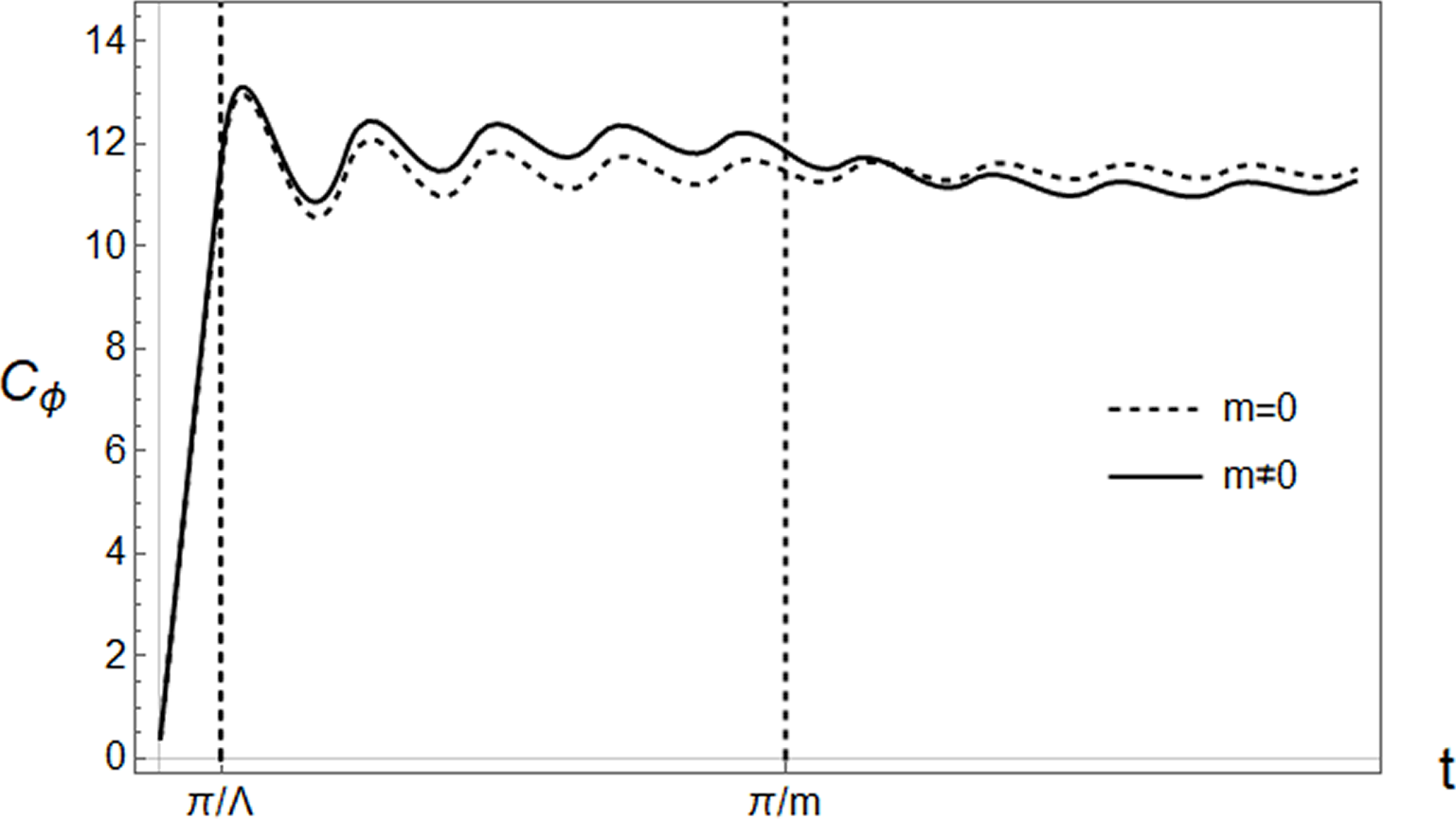}
		\caption{}
		\label{subfig:1dScalar}
	\end{subfigure} \hfill
	\begin{subfigure}[b]{0.49\textwidth}
	\centering \includegraphics[width=\textwidth]{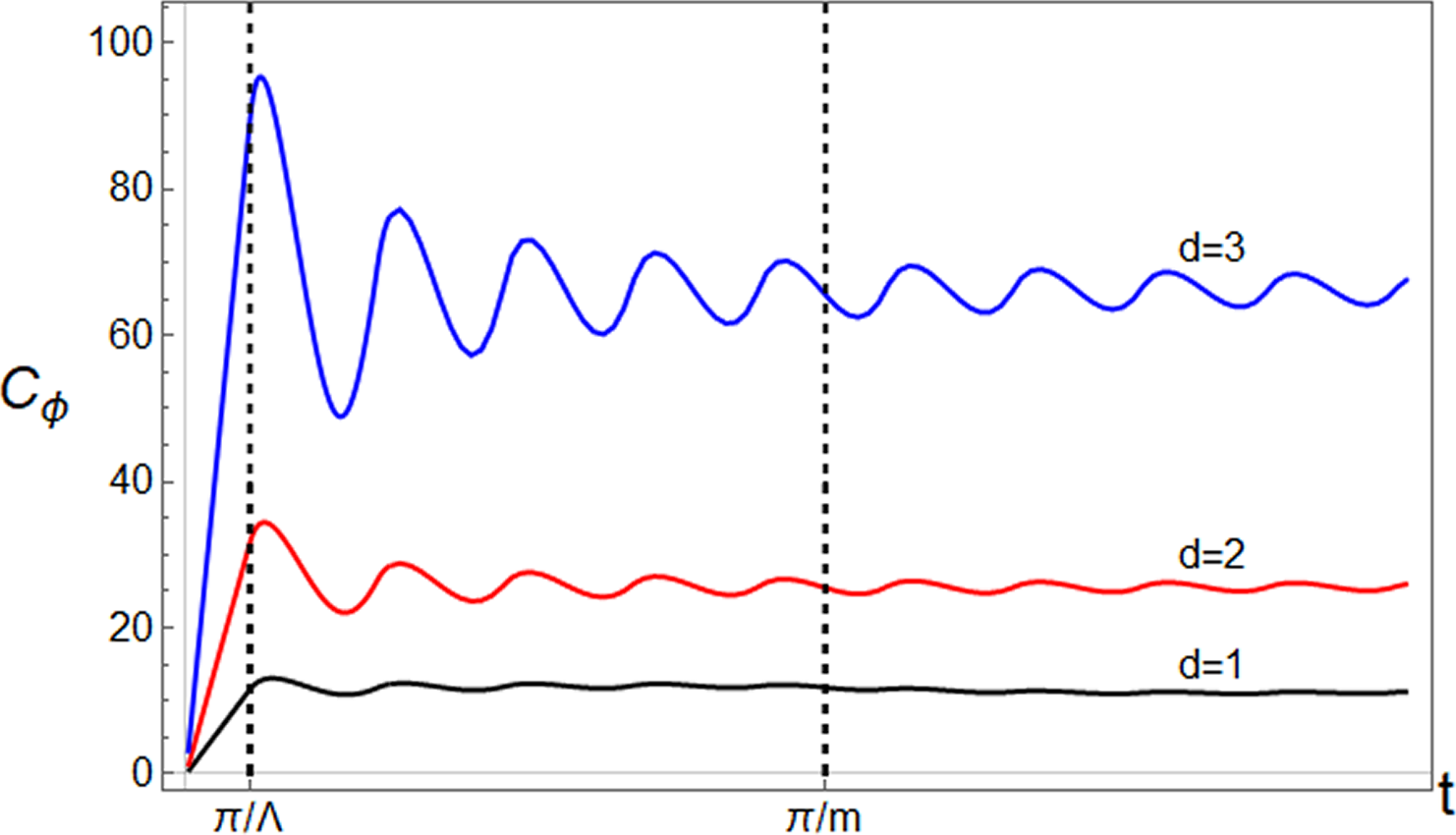}
	\caption{}
	\label{subfig:ScalarCombined}
	\end{subfigure}
	\caption{The operator complexity (\ref{eq:scalarFieldComplexityPhi}) of the time evolution operator for a free scalar field of mass $m$ with UV cutoff $\Lambda$ (here chosen as $m = \Lambda/10$) shows a sharp growth ${\mathcal C}_{\phi}\sim 2\Lambda t$ at early times, some transient behavior at intermediate times $\pi/\Lambda < t < \pi/m$, and a saturation (with damped oscillations) at late times $t > \pi/m$. (a) The transient intermediate behavior is shown for $d=1$ for $m\neq 0$ (solid); when $m=0$ (dashed) there is no transient period, and the operator complexity saturates after its initial growth. (b) The growth and saturation of the operator complexity is similar across different dimensions.}
	\label{fig:ScalarComplexity}
\end{figure}

At intermediate times $\pi/\Lambda \ll t \ll \pi/m$, the complexity for the UV modes $p^2 \gg m^2$ are rapidly oscillating, but the complexity for the IR modes $p^2 < m^2$ are still increasing with time, $v_3^{IR} \sim 2 m t$. Since the UV modes dominate the integral, this leads to a transient period of mild time dependence, as seen for $d=1$ in Figure \ref{subfig:1dScalar}.
%By separating the integral into separate integrals over the IR and UV modes, we find that the resulting total complexity of the scalar field is weakly growing with time,
%\be
%{\mathcal C}_{\phi} \sim \frac{L^{d/2}}{\pi^{d/2}} \sqrt{\int_{\rm IR} m^2 t^2\, d^dp + \int_{\rm UV} \left(\frac{\pi}{2}\right)^2\, d^dp} \sim L^{d/2} \sqrt{\Lambda^d + m^{d+2} t}\, .
%\ee
Finally, at late times $t \gg \pi/m$, the complexity of each mode is rapidly oscillating with an ${\mathcal O}(1)$ value, so the total complexity of the scalar field roughly saturates (up to oscillations that will damp out over time) to an average value set by the UV scale,
\be
{\mathcal C}_{\phi} \sim \frac{L^{d/2}}{\pi^{d/2}} \sqrt{\int^\Lambda d^d p} \sim L^{d/2} \Lambda^{d/2}\, .
\label{eq:phiLateComplexity}
\ee
Putting this behavior together, we get a plot of complexity versus time Figure \ref{fig:ScalarComplexity} that has a sharp linear rise followed by a saturation with damped oscillations, consistent across multiple values of the dimension $d$ as shown in Figure \ref{subfig:ScalarCombined}.
Taking the scalar field to be massless $m\rightarrow 0$ removes the intermediate transient behavior, so that the complexity simply saturates after the sharp rise at early times, as seen in Figure \ref{subfig:1dScalar}.

The volume and UV divergences in the saturated operator complexity (\ref{eq:phiLateComplexity}) of the free scalar field Hamiltonian evolution operator are similar to the divergences found in \cite{Jefferson} for the state complexity of the ground state of a free scalar field in $d$-dimensions.
Naturally, the power of $1/2$ in (\ref{eq:phiLateComplexity}) is coming from the square root in the geometric length of the circuit depth (\ref{eq:CircuitDepth}). 
The Euler-Arnold minimization procedure simultaneously minimizes the ``squared length" circuit depth
\be
\tilde{\mathcal C} = \min_{\{V^I\}} \tilde{\mathcal D}\left[V^I\right] = \min_{\{V^I\}} \int_0^1 G_{IJ}V^I V^J\ ds\, ,
\ee
so the corresponding saturated complexity for the scalar field would scale with the volume and UV cutoff instead as
\be
\tilde {\mathcal C}_{\phi} \sim L^d \Lambda^d\, .
\ee
This divergence structure compares more favorably with that of common proposals of complexity from holography \cite{Reynolds:2016rvl,Carmi:2016wjl}, as well as other studies of scalar field complexity \cite{Jefferson}.

\subsection{Quantum Scalar Cosmological Perturbations}

Another interesting application of operator complexity to field theory is its application to the unitary evolution of quantum cosmological perturbations in an expanding background. We will briefly review the description of quantum cosmological perturbations as squeezed states; see \cite{Mukhanov,Grishchuk,Albrecht,Martin1,Martin2} for details.
We will work in $(3+1)$-dimensions and take a spatially flat Friedmann-Lemaitre-Robertson-Walker (FLRW) metric
\be
ds^2 = -dt^2 + a(t)^2 d\vec{x}^2 = a(\eta)^2 \left(-d\eta^2 + d\vec{x}^2\right)
\ee
where $\eta$ is known as the conformal time. The Hubble expansion rate of this background is characterized by the time derivative of the scale factor $H = \dot a/a = a'/a^2$, where a dot denotes a derivative with respect to cosmic time $t$ and a prime denotes a derivative with respect to conformal time $\eta$.
On this background, we will consider fluctuations of a light scalar field $\phi = \phi_0(t) + \delta \phi(x,t)$; the fluctuations $\delta \phi$ combine with linearized fluctuations of the metric to form gauge-invariant perturbations, such as the curvature perturbation ${\mathcal R}$.
When written in terms of the Mukhanov-Sasaki variable $v \equiv z {\mathcal R}$, where $z \equiv a \sqrt{2\epsilon}$ with $\epsilon = -\dot H/H^2$, the action expanded to quadratic order becomes
\be
S = \frac{1}{2} \int d\eta\, d^3x \left[v'^2 - (\vec{\nabla} v)^2 + \left(\frac{z'}{z}\right) v^2 - 2 \frac{z'}{z}\, v' v\right]\, .
\label{eq:CosmoPertAction}
\ee
The action (\ref{eq:CosmoPertAction}) represents a massless scalar field coupled to an external time-dependent source due to the expanding cosmological background.

According to the inflationary model of the early universe, the observed classical perturbations in the early universe began as quantum cosmological perturbations stretched to large scales by the rapid expansion of inflation \cite{Mukhanov}.
More generally, quantum fields on time-dependent and curved backgrounds can lead to interesting effects such as particle production and entanglement.
We will promote our scalar cosmological perturbation to a quantum field and expand in (continuous) Fourier modes 
%as in (\ref{eq:scalarFourier}) (we will consider the box to be much larger than the corresponding Hubble length $L \gg H^{-1}$, and again anticipate taking $L\rightarrow \infty$ at the end of the calculation), 
leading to the Hamiltonian
\be
\hat H_{\rm cosmo} = \int d^3k\, \hat {\mathcal H}_{\vec{k}} = \int d^3k\left[ k \left(\hat a^\dagger_{\vec{k}} \hat a_{\vec{k}} + \hat a_{-\vec{k}} \hat a^\dagger_{-\vec{k}}\right) - i \frac{z'}{z} \left(\hat a_{\vec{k}} \hat a_{-\vec{k}} - \hat a^\dagger_{\vec{k}} \hat a^\dagger_{-\vec{k}} \right)\right]
\label{eq:CosmoH}
\ee
where we wrote the creation and annihilation operators with explicit factors of $\vec{k}_{\vec{n}} = \vec{n} \pi/L$ to illustrate that the first term represents the free Hamiltonian evolution of the modes $\pm \vec{k}_{\vec{n}}$, while the second term represents particle creation from the time-dependent cosmological background, entangling modes of opposite momenta.

Let us first consider the unitary evolution associated with the evolution of a single $(\vec{k},-\vec{k})$ pair of modes; we will then return to considering the total evolution of all of the modes.
The unitary time-evolution operator
\be
\hat U_{\vec{k}}(\eta) = {\mathcal P}\, e^{-i \int_0^\eta \hat {\mathcal H}_{\vec{k}}(\tilde \eta) d\tilde\eta}
\ee
can be rewritten in terms of two-mode squeezing and rotation operators
\be
\hat U_{\vec{k}}(\eta) =  \hat {\mathcal S}_{\vec{k}}\left(r_k, \phi_k\right) \hat {\mathcal R}_{\vec{k}}\left(\theta_k\right)
\ee
where
\be
\hat {\mathcal S}_{\vec{k}} &=& \mbox{exp}\left[\frac{r_k(\eta)}{2}\left(e^{-2i\phi_k(\eta)}\hat a_{\vec{k}}\hat a_{-\vec{k}}-e^{2i\phi_k}\hat a^\dagger_{\vec{k}}\hat a^\dagger_{-\vec{k}}\right)\right] \label{eq:CosmoSqueezeDef}\\
\hat {\mathcal R}_{\vec{k}} &=& \mbox{exp}\left[-i\theta_k(\eta)\left(\hat a_{\vec{k}} \hat a^\dagger_{\vec{k}} + \hat a^\dagger_{-\vec{k}}\hat a_{-\vec{k}}\right)\right]\label{eq:CosmoRotDef}
\ee
with $k = |\vec{k}|$, are similar to the single-mode squeezing and rotation operators (\ref{eq:SqueezeDef}) of Section \ref{sec:OP_SU11}.
Indeed, the two-mode operators (\ref{eq:CosmoSqueezeDef},\ref{eq:CosmoRotDef}) can be built from the fundamental set of operators
\be
\hat e_1^{\vec{k}} = \frac{\hat a_{\vec{k}} \hat a_{-\vec{k}} + \hat a^\dagger_{\vec{k}} \hat a^\dagger_{-\vec{k}}}{2}, \hspace{.2in}
\hat e_2^{\vec{k}} = \frac{i}2\left(\hat a_{\vec{k}} \hat a_{-\vec{k}} - \hat a^\dagger_{\vec{k}} \hat a^\dagger_{-\vec{k}}\right), \hspace{.2in}
\hat e_3^{\vec{k}} = \frac{\hat a_{\vec{k}}\hat a^\dagger_{\vec{k}} + \hat a^\dagger_{-\vec{k}}\hat a_{-\vec{k}}}{2}
\ee
which satisfy the $\mathfrak{su(1,1)}$ algebra (\ref{su11algebra}). This allows us to use the calculations of Section \ref{sec:SqueezeRotComplexity} so that the complexity for the individual pair of modes $(\vec{k},-\vec{k})$ is
\be
{\mathcal C}_{\vec{k}}(\eta) = \sqrt{4r_k(\eta)^2 + v_3(\eta)^2}
\label{eq:CosmoPertkComplexity}
\ee
where $v_3(\eta)$ is given in terms of $\theta_k(\eta)$ by (\ref{eq:SU11BoundarySolutionApprox}).

Before using the time-dependence of the squeeze and rotation parameters to determine the resulting time-dependence of the complexity (\ref{eq:CosmoPertkComplexity}), we immediately notice that by construction, the complexity (\ref{eq:CosmoPertkComplexity}) does not include any information about the reference state. 
In particular, we did not need to assume that the quantum state is in its ground state in the asymptotic past, as is typically done to study the time-evolution of the quantum state of cosmological perturbations.
Our complexity of the time-evolution operator of the cosmological background (\ref{eq:CosmoPertkComplexity}) is thus insensitive to details such as the initial state of the cosmological perturbations, in contrast to state-dependent methods such as \cite{cosmology1,cosmology2,Lehners_2021,Adhikari:2021ked}.
Thus, it is potentially useful as a measure of the complexity of the background, independent of the details of state preparation.

\begin{figure}[t]
	\centering
	\begin{subfigure}[b]{0.49\textwidth}
		\centering\includegraphics[width=\textwidth]{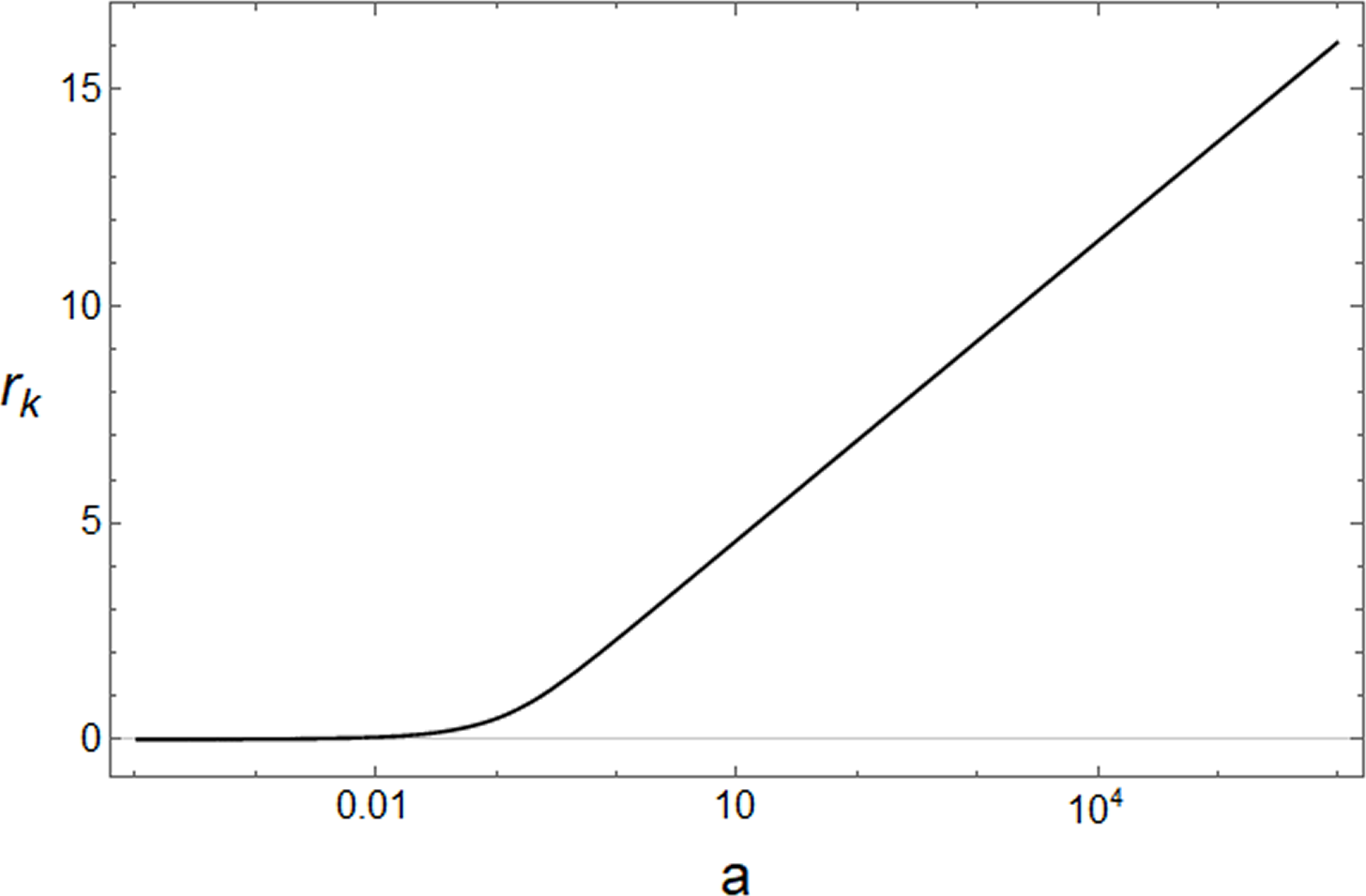}
		\caption{}
		\label{subfig:rplot}
	\end{subfigure}\hfill
	\begin{subfigure}[b]{0.49\textwidth}
		\centering \includegraphics[width=\textwidth]{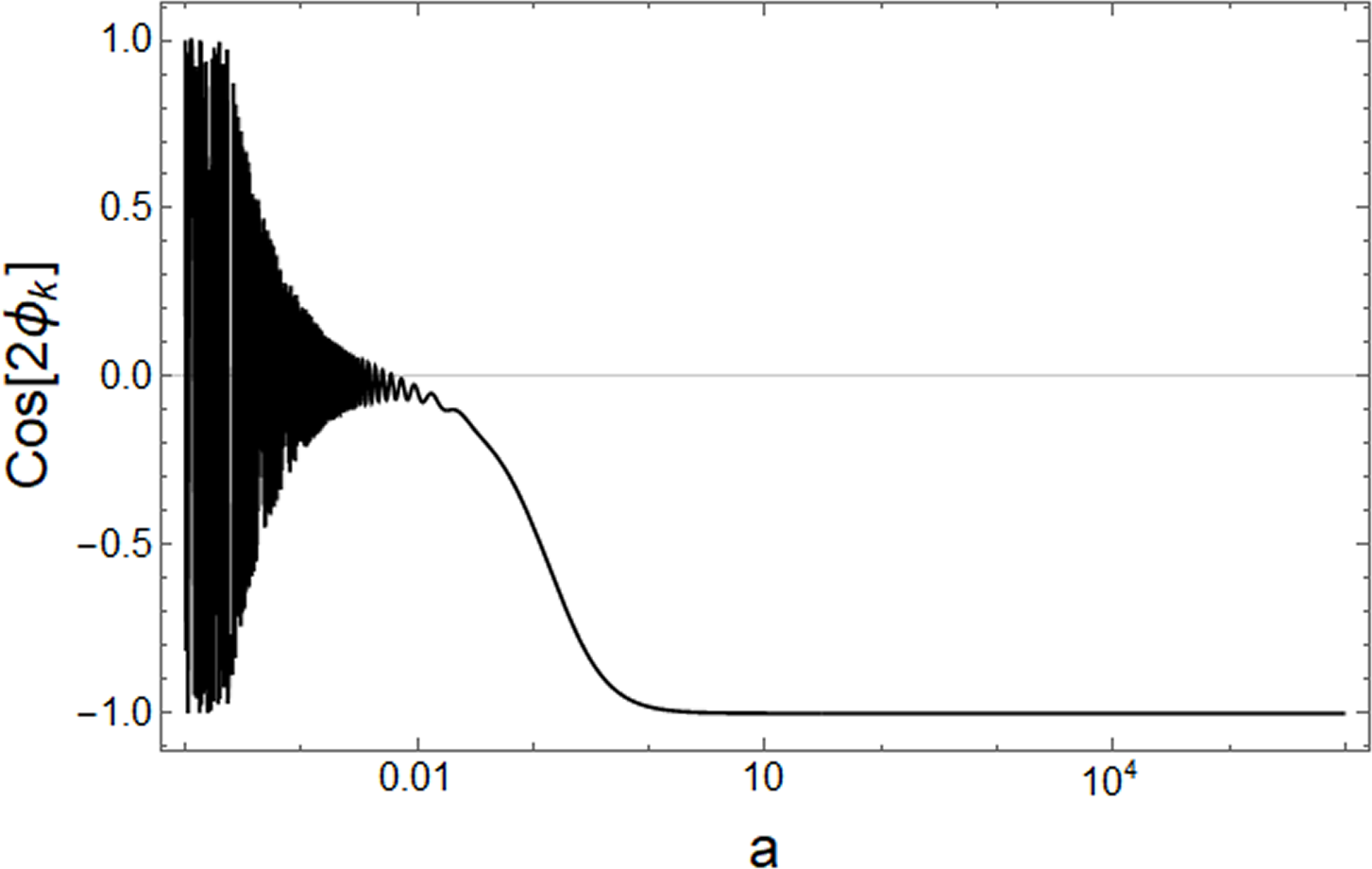}
		\caption{}
		\label{subfig:phiplot}
	\end{subfigure}
	\begin{subfigure}{0.49\textwidth}
			\centering \includegraphics[width=\textwidth]{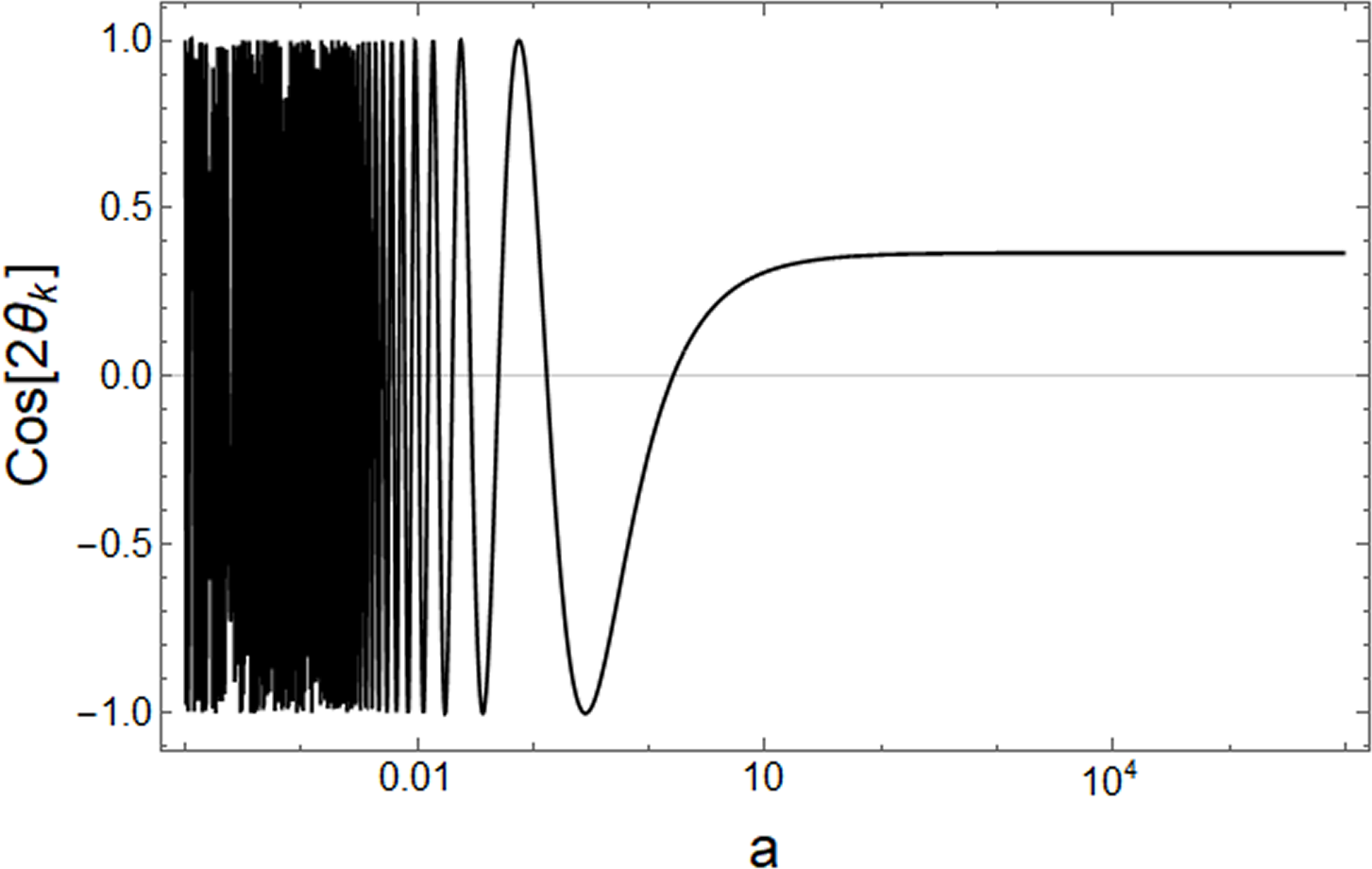}
		\caption{}
		\label{fig:thetaplot}
	\end{subfigure}
	\caption{(a) Squeezing parameter $r_k$, (b) squeezing angle $\phi_k$, and (c) rotation angle $\theta_k$, each as a function of the scale factor $a = a(\eta)$ for a de Sitter background and a fixed co-moving wavenumber $k$. While the mode is inside the horizon $k H_{dS} \gg a$, the squeezing is small $r_k \ll 1$ and the squeezing and rotation angles vary with time. When the mode exits the horizon, the squeezing begins to grow $r_k \sim \ln a$ and the squeezing angle and rotation angles ``freeze-out'' to fixed values.}
	\label{fig:SqueezingSolutions}
\end{figure}

%\begin{figure}[h]
%	\centering \includegraphics[width=0.49\textwidth]{thetaplot}
%	\caption{}
%	\label{fig:thetaplot}
%\end{figure}

The time-dependence for the squeezing parameter $r_k(\eta)$ squeezing angle $\phi_k(\eta)$ and rotation angle $\theta_k(\eta)$ arises from the Heisenberg equations of motion with the Hamiltonian $\hat {\mathcal H}_{\vec{k}_{\vec{n}}}$ (e.g. see \cite{Albrecht,Martin1,Martin2})
\be
\frac{dr_k}{d\eta} &=& -\frac{z'}{z} \cos(2\phi_k)\, ; \label{eq:rkEOM} \\
\frac{d\phi_k}{d\eta} &=& k + \frac{z'}{z}\coth(2r_k) \sin(2\phi_k)\, ; \label{eq:phikEOM} \\
\frac{d\theta_k}{d\eta} &=& k - \frac{z'}{z} \tanh(r_k) \sin(2\phi_k)\, . \label{eq:thetakEOM}
\ee
An exact solution of particular interest is known for de Sitter space, $a(\eta) = -1/(H_{dS} \eta)$ for $\eta < 0$
\be
r_k &=& \sinh^{-1} \left(\frac{1}{2|k \eta|}\right) = \sinh^{-1} \left(\frac{a}{2kH_{dS}}\right) \label{eq:rkdS} \\
\phi_k &=& -\frac{\pi}{4} - \frac{1}{2} \tan^{-1}\left(\frac{1}{2|k\eta|}\right) = -\frac{\pi}{4} - \frac{1}{2} \tan^{-1}\left(\frac{a}{2kH_{dS}}\right) \label{eq:phikdS} \\
\theta_k &=& |k\eta| - \tan^{-1}\left(\frac{1}{2|k\eta|}\right) = \frac{k H_{dS}}{a} - \tan^{-1} \left(\frac{a}{2 k H_{dS}}\right) \label{eq:thetakdS}
\ee
At early times $\eta \rightarrow -\infty$ (or $k H_{dS} \gg a$), the mode is inside the horizon and the squeezing $r_k \ll 1$ is small, while $\phi_k, \theta_k$ are varying with time.
As the mode exits the horizon at late times $|k \eta| \ll 1$ ($a \gg k H_{dS}$), the squeezing begins to grow $r_k \sim \ln a$ and the squeezing and rotation angles ``freeze out'' to their superhorizon values.
This behavior is shown numerically in Figures \ref{fig:SqueezingSolutions} and \ref{fig:thetaplot}.

\begin{figure}[t]
\centering \includegraphics[width=.9\textwidth]{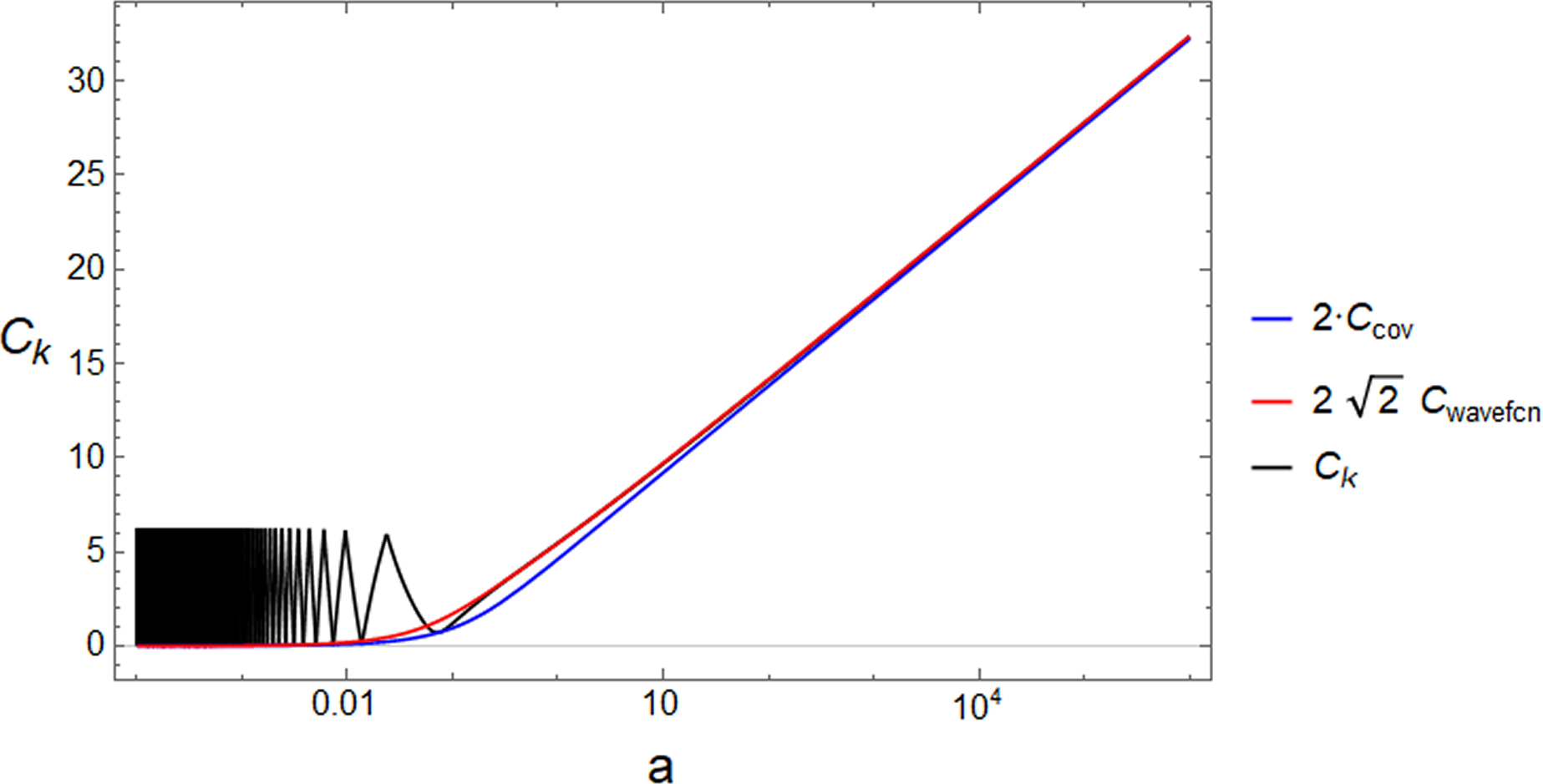}
\caption{The complexity of a single $(\vec{k},-\vec{k})$ mode pair of cosmological perturbations in a de Sitter background for several different measures of complexity as a function of the scale factor. Operator complexity ${\mathcal C}_{\vec{k}}$ (\ref{eq:CosmoPertkComplexity}) (black), studied in this paper, is sensitive to the oscillating phase $\theta_k$ at early times. The covariance matrix complexity of the corresponding state ${\mathcal C}_{\rm cov}$ (\ref{eq:CosmoCovkComplexity}) (blue) is directly proportional to the squeezing parameter $r_k$, which is insensitive to the phase. The wavefunction-method complexity of the corresponding state ${\mathcal C}_{\rm wavefcn}$ (\ref{eq:CosmoWaveFcnkComplexity}) (red) is also insensitive to the phase, though its sensitivity to the squeezing angle gives very slight differences in intermediate growth. All of these different measures of complexity grow linearly with the number of e-folds $N_e\sim \ln a$ at late times (with slightly different numerical coefficients due to differing conventions).}
\label{fig:CosmoComplexity}
\end{figure}

For a fixed time, small wavelength modes $|k \eta| \gg 1$ have small squeezing $r_k \ll 1$ so the
the complexity (\ref{eq:CosmoPertkComplexity}) is dominated by the rapidly oscillating contribution from the phase $\theta_k$ through $v_3$, and the complexity is ${\mathcal C}_{\vec{k}} \sim {\mathcal O}(1)$. For long-wavelength modes $|k \eta| \ll 1$, the mode is outside the Hubble horizon and the complexity becomes dominated by the squeezing parameter
\be
{\mathcal C}_{\vec{k}}(\eta) \sim 2r_k \sim 2\ln\left(\frac{1}{2|k\eta|}\right) \sim 2\ln\left(a(\eta)/a_e\right) \sim 2 N_e
\ee
where $a_e = -1/(H_{dS} k)$ is the scale factor at horizon exit, and in the last step we wrote the result in terms of the number of e-folds of expansion since the horizon exit of the mode. A plot of the full numerical solution of the single-mode complexity (\ref{eq:CosmoPertkComplexity}) as a function of the scale factor of the universe illustrating this behavior can be found in Figure \ref{fig:CosmoComplexity}.

It is interesting to compare the operator complexity (\ref{eq:CosmoPertkComplexity}) for the cosmological perturbation $(\vec{k},-\vec{k})$ mode pair to the corresponding state complexity.
There are several different techniques for calculating state complexity\footnote{For the purposes of comparison, each of these techniques assumes a geometric cost function and a Bunch-Davies vacuum state as a reference state.}, including the wavefunction method (see \cite{Jefferson, cosmology1, cosmology2}) with corresponding complexity
%, calculated in \cite{cosmology1} (see also \cite{cosmology2,Lehners_2021} {\bf [Satayan]})
\be
{\mathcal C}_{\rm wavefcn} = \frac{1}{\sqrt{2}} \sqrt{\left| \ln \left|\frac{1+e^{-2i\phi_k}\, \tanh r_k}{1-e^{-2i\phi_k}\, \tanh r_k}\right|\right|^2 + \rm{arctan}\left(2\sin 2\phi_k\, \sinh r_k\, \cosh r_k\right)^2}\, , \label{eq:CosmoWaveFcnkComplexity}
\ee
and the covariance matrix method (for examples, see the appendix of \cite{cosmology1} as well as \cite{Lehners_2021}), with corresponding complexity,
\be
{\mathcal C}_{\rm cov} = r_k\, . \label{eq:CosmoCovkComplexity}
\ee
In the long-wavelength (superhorizon) limit $k H_{dS} \ll a$, all of the measures of complexity (\ref{eq:CosmoPertkComplexity},\ref{eq:CosmoWaveFcnkComplexity},\ref{eq:CosmoCovkComplexity})  
grow in a similar way as the number of e-folds\footnote{The wavefunction complexity is smaller by a factor of $2\sqrt{2}$ due to the conventions of (\ref{eq:CosmoWaveFcnkComplexity}); the covariance matrix method complexity is similarly different by a factor of 2 due to conventions.}
\be
{\mathcal C}_{\vec{k}} \sim 2\sqrt{2}\,{\mathcal C}_{\rm wavefcn} \sim 2 {\mathcal C}_{\rm cov}\sim 2\ln\left(a(\eta)/a_e\right)\sim 2N_e\, .
\ee
%More generally, the operator complexity and covariance matrix complexity are in fact nearly identical up to a factor of two, except for the presence of $v_3^2$ in (\ref{eq:CosmoPertkComplexity}) due to the sensitivity of operator complexity to the phase of the operator.
The difference between the measures of complexity is primarily apparent at small wavelengths $a \ll k H_{dS}$ when the mode is within the horizon: the operator complexity (\ref{eq:CosmoPertkComplexity}) rapidly oscillates between $0$ and $2\pi$, while both of the state complexities (\ref{eq:CosmoWaveFcnkComplexity}), (\ref{eq:CosmoCovkComplexity}) smoothly go to zero in this UV limit. This differing behavior of the complexities can be seen in Figure \ref{fig:CosmoComplexity} in the limit of small scale factor $a$, which shows each of these complexities for a fixed wavelength $k$.

Now that we have examined the operator complexity for a single $(\vec{k},-\vec{k})$ mode of cosmological perturbations, let us now consider its extension to field theory by including an integral over the Fourier modes of the Hamiltonian (\ref{eq:CosmoH}).
Analogously to (\ref{eq:scalarfieldComplexity}), the total cosmological complexity (\ref{eq:CosmoPertkComplexity}) now becomes
\be
{\mathcal C}_{\rm op}^{\rm tot} \rightarrow L^{3/2} \sqrt{\int^\Lambda \left(4r_k(\eta)^2 + v_3(\eta)^2\right)d^3k}\, .
\label{eq:CosmoTotComplexity}
\ee
As noted above, in the UV limit $|k\eta| \gg 1$ the integrand is dominated by the rapid oscillations of the phase about an ${\mathcal O}(1)$ value.
In the IR limit $|k\eta| \ll 1$, the integrand is dominated by the squeezing parameter $r_k(\eta)\sim \ln\left(\frac{1}{|k\eta|}\right) \gg 1$.
The integral in (\ref{eq:CosmoTotComplexity}) thus roughly splits into two regimes, which combine as
\be
{\mathcal C}_{\rm op}^{\rm tot} &\sim& L^{3/2} \sqrt{\int_{IR} 4 r_k(\eta)^2\, d^3 k + \int_{UV} v_3(\eta)^2\, d^3 k}
\sim L^{3/2} \sqrt{\int_0^{|\eta|^{-1}} k^2 \ln^2\left(\frac{1}{|k\eta|}\right) dk + \int_{|\eta|^{-1}}^\Lambda k^2 dk} \nonumber \\
	&\sim& L^{3/2} \sqrt{\frac{\alpha_1}{|\eta|^3} + \alpha_2 \Lambda^3}\, ,
\label{eq:CosmoTotComplexityIRUV}
\ee
up to ${\mathcal O}(1)$ factors absorbed into the $\alpha_i$ constants.
As expected from Section \ref{subsec:FreeField}, the UV part of the integral diverges as $\Lambda^3$ and is independent of the de Sitter background. Surprisingly, this competes with the first term, which represents growth in complexity in the IR due to the squeezing and is proportional to the growth in volume of de Sitter space $1/|\eta|^3 \sim H_{dS}^3 a(\eta)^3 \sim H_{dS}^3 e^{N_e}$. In fact, for a UV cutoff near the Planck scale $\Lambda \sim 10^{19}$ GeV and 60 e-folds of expansion, the squeezing contribution to the complexity dominates over that from the UV modes for Hubble scales down to approximately $H_{dS} > 1$ keV!

How does this result for the total {\it operator} complexity of quantum cosmological perturbations (\ref{eq:CosmoTotComplexityIRUV})
compare to the total state complexities (\ref{eq:CosmoCovkComplexity}),(\ref{eq:CosmoWaveFcnkComplexity}) in the field theory limit?
The generalization of the covariance matrix-method state complexity (\ref{eq:CosmoCovkComplexity}) to include multiple modes in the continuum limit should be the geometric sum
\be
{\mathcal C}_{\rm cov}^{\rm tot} \sim L^{3/2} \sqrt{\int^\Lambda r_k^2\, d^3k}\, .
\label{eq:CosmoCovTotComplexity1}
\ee
The IR contribution to the integral (\ref{eq:CosmoCovTotComplexity1}) is the same as for (\ref{eq:CosmoTotComplexityIRUV}), while in the UV we have $r_k \approx 1/(2k|\eta|) \ll 1$ so (\ref{eq:CosmoCovTotComplexity1}) becomes
\be
{\mathcal C}_{\rm cov}^{\rm tot} \sim L^{3/2} \sqrt{\int_{IR} r_k^2\, d^3k + \int_{UV} r_k^2\, d^3k} \sim L^{3/2} \sqrt{\frac{\beta_1}{|\eta|^3} + \beta_2 \frac{\Lambda}{|\eta|^2}}\, ,
\label{eq:CosmoCovTotComplexity2}
\ee
where $\beta_i \sim {\mathcal O}(1)$. Interestingly, the UV divergence is weaker here, depends on the de Sitter scale through $\Lambda/|\eta|^2 \sim \Lambda H_{dS}^2 a(\eta)^2$, and scales with the square of the scale factor.

Finally, let's take the continuum field theory limit of the wavefunction-method complexity (\ref{eq:CosmoWaveFcnkComplexity})
\be
{\mathcal C}_{\rm wavefcn}^{\rm tot} &\sim & \frac{L^{3/2}}{\sqrt{2}} \sqrt{\int^\Lambda \left(\left| \ln \left|\frac{1+e^{-2i\phi_k}\, \tanh r_k}{1-e^{-2i\phi_k}\, \tanh r_k}\right|\right|^2 + \rm{arctan}\left(2\sin 2\phi_k\, \sinh r_k\, \cosh r_k\right)^2\right) d^3k} \nonumber \\
&& \,
\label{eq:CosmoWaveFcnTotComplexity1}
\ee
Again, we can approximate this expression by dividing the integral into IR $|k\eta| \ll 1$ and UV $|k\eta| \gg 1$ parts, $\int^\Lambda \approx \int_{IR} + \int_{UV}$.
In the IR, we take the approximate solutions (see \cite{cosmology2}) $r_k \approx \ln(1/|k\eta|) \gg 1, \phi_k \approx -\pi/2$ and the IR integral is dominated by the first term of the integrand
\be
\int_0^{|\eta|^{-1}} \left| \ln \left|\frac{1+e^{-2i\phi_k}\, \tanh r_k}{1-e^{-2i\phi_k}\, \tanh r_k}\right|\right|^2\,k^2\,  dk 
	\sim \int_0^{|\eta|^{-1}} k^2 \ln^2\left(\frac{1}{|k\eta|}\right)\, dk \sim \lambda_1 \frac{1}{|\eta|^3}
	\label{eq:WaveFcnComplexityIRLimit}
\ee
where $\lambda_1\sim {\mathcal O}(1)$ is an order one constant. As with the IR contributions of the other measures of complexity, this scales as the volume growth of de Sitter, exponentially with the number of e-folds $1/|\eta|^3 = H_{dS}^3 a(\eta)^3 \sim H_{dS}^3 e^{3N_e}$.
For the UV $|k\eta| \gg 1$, the squeezing parameter and angle have the approximate solutions \cite{cosmology2} $r_k \approx 1/(2|k\eta|) \ll 1, \phi_k \approx -\pi/4 - 1/(4|k\eta|)$, where we kept the subleading dependence in $\phi_k$.
Inserting these solutions into the integrand of (\ref{eq:CosmoWaveFcnTotComplexity1}), the leading behavior in the UV $|\eta|^{-1} \leq k \leq \Lambda$ is
\be
&&\int_{|\eta|^{-1}}^\Lambda k^2 \left(\left| \ln \left|\frac{1+e^{-2i\phi_k}\, \tanh r_k}{1-e^{-2i\phi_k}\, \tanh r_k}\right|\right|^2 + \rm{arctan}\left(2\sin 2\phi_k\, \sinh r_k\, \cosh r_k\right)^2\right)\, dk \nonumber \\
\approx && \int_{|\eta|^{-1}}^\Lambda k^2 \left(\frac{8}{(k|\eta|)^4}\right)\, dk \approx \frac{\lambda_2 }{|\eta|^4}\left(|\eta|-\frac{1}{\Lambda}\right) \rightarrow \frac{\lambda_2}{|\eta|^3}\, ,
\label{eq:WaveFcnComplexityUVLimit}
\ee
with $\lambda_2 \sim {\mathcal O}(1)$.
Remarkably, the UV integral (\ref{eq:WaveFcnComplexityUVLimit}) does not diverge, and we can take $\Lambda \rightarrow 0$ to obtain a UV-finite result, as we did in the last step above. We then find that the UV contribution (\ref{eq:WaveFcnComplexityUVLimit}) also scales as $1/|\eta|^3$, similar to the IR part of the integral.
Combining the IR (\ref{eq:WaveFcnComplexityIRLimit}) and UV (\ref{eq:WaveFcnComplexityUVLimit}) contributions to the integrand of (\ref{eq:CosmoWaveFcnTotComplexity1}), then, we obtain a result for the total field theory state complexity (wavefunction method) of cosmological perturbations in de Sitter space as
\be
{\mathcal C}_{\rm wavefcn}^{\rm tot} \sim \gamma L^{3/2} \frac{1}{|\eta|^{3/2}} \sim \gamma L^{3/2} H_{dS}^{3/2} a(\eta)^{3/2}\, ,
\label{eq:CosmoWaveFcnTotComplexity2}
\ee
where again $\gamma$ is some other  ${\mathcal O}(1)$ constant.

Summarizing our results for the three different complexities 
(\ref{eq:CosmoTotComplexityIRUV}),(\ref{eq:CosmoCovTotComplexity2}),(\ref{eq:CosmoWaveFcnTotComplexity2})
of cosmological perturbations in the field theory limit,
\begin{align}
{\mathcal C}_{\rm op}^{\rm tot} \sim & L^{3/2} \sqrt{\frac{\alpha_1}{|\eta|^3} + \alpha_2 \Lambda^3} \hspace{.1in} &\mbox{Operator Complexity} \label{eq:CosmoOpTotComplexity}\\
{\mathcal C}_{\rm cov}^{\rm tot} \sim & L^{3/2} \sqrt{\frac{\beta_1}{|\eta|^3} + \beta_2\frac{\Lambda}{|\eta|^2}} & \mbox{Covariance Matrix-Method State Complexity} \label{eq:CosmoCovTotComplexity3}\\
{\mathcal C}_{\rm wavefcn}^{\rm tot} \sim & \gamma L^{3/2} \frac{1}{|\eta|^{3/2}}  & \mbox{Wavefunction-Method State Complexity} \label{eq:CosmoWaveFcnTotComplexity3}
\end{align}

The wavefunction-method complexity (\ref{eq:CosmoWaveFcnTotComplexity3}) simply scales as the square root of the volume, and is UV finite in contrast to the operator complexity (\ref{eq:CosmoOpTotComplexity}) or covariance matrix-method complexity (\ref{eq:CosmoCovTotComplexity3}).
Nevertheless, as we argued above, even for a Planck-scale UV cutoff the IR terms dominate for around 60 e-folds of expansion when $H_{dS} > 1$ keV.
Assuming the IR term dominates in all of the complexities above, the scaling ${\mathcal C}^{\rm tot}\sim a(\eta)^{3/2}$ means that during inflation, the total complexity of the universe due to the cosmological perturbations grows by a factor,
\be
\frac{{\mathcal C}^{\rm tot}(\eta_f)}{{\mathcal C}^{\rm tot}(\eta_i)} \sim \left(\frac{a_f}{a_i}\right)^{3/2} \sim e^{3N_e/2}\, ,
\ee
which is a factor of $\sim e^{90} \sim 10^{39}$ for $N_e \sim 60$ e-folds of inflation.

%As we saw in the previous subsection, the operator complexity associated with the phase angle $\theta_k$ will give rise to a UV-divergence when we consider the total field-theory complexity, while the corresponding state complexity (\ref{eq:CosmoStatekComplexity}) will be regular in the UV.

%%%%%%%%%%%%%%%%%%%%%%%%%%%%%%%%%%%%%%%%%%%%%%%%%%%%%%%%%%%%%
%%%%%%%%%%%%%%%%%%%%%%%%%%%%%%%%%%%%%%%%%%%%%%%%%%%%%%%%%%%%%
%%%%%%%%%%%%%%%%%%%%%%%%%%%%%%%%%%%%%%%%%%%%%%%%%%%%%%%%%%%%%
\section{Discussion}
\label{sec:Discussion}

As a step towards a better understanding of the complexity of operators in scalar field quantum field theories, we analyzed the operator complexity associated with 
the displacement, squeeze, and rotation operators of a quantum harmonic oscillator.
Applying the approach of \cite{Balasubramanian:2019wgd,Balasubramanian:2021mxo},
we define the complexity of a target operator as a minimal length geodesic between two points, identified with the identity and the target operator, in a geometry associated with the group algebra of the fundamental operators that generate the operator.
As discussed in the Introduction, this notion of operator complexity is independent of the choice of reference and target states.

We focused here on two sets of unitary operators of the quantum harmonic oscillator -- the displacement operator, and the squeeze and rotation operators -- 
which together characterize any unitary operation that is at most quadratic in the creation and annihilation operators.
The displacement operator can be constructed with generators of the Heisenberg group, and the resulting operator-space geometry is 3-dimensional hyperbolic space.
The corresponding complexity of a (time-dependent) displacement operator is proportional to the magnitude of the coherent state parameter, independent of time. 
This result is similar to some previous results for the complexity of coherent states obtained using state-based complexity techniques \cite{Guo:2018kzl,Guo:2020dsi}, but is in contrast to other results for the complexity of the displacement operator in which a non-trivial dependence on time was found \cite{Bhattacharyya:2020art,Caputa:2021sib}. It would be interesting to study the reasons for these differences.

We also considered the squeeze and rotation operators, which are elements of the group SU(1,1) (isomorphic to ${\rm SL}(2,\mathbb{R})$), finding again a non-compact, negatively curved group manifold.
As two simple examples, we confirmed that the complexity for time evolution by a free harmonic oscillator is bounded by $2\pi$ and oscillates with a frequency given by the quantum revival time, while the complexity for an inverted harmonic oscillator grows linearly with time without bound, reflecting the instability of the inverted oscillator.
More generally, a generic quadratic Hamiltonian can be written as the product of squeeze and rotation operators, and the corresponding operator grows  linearly in the squeezing parameter for large squeezing, and is independent of the squeezing angle.
This stands in contrast with the wavefunction-method complexity of a squeezed state \cite{cosmology1,cosmology2}, in which the squeezing angle plays an important role in determining the time-dependent behavior of complexity.
It would be interesting to study further the relationships and qualitative differences between the wavefunction-based and operator-based methods for calculating complexity.

Our analysis of the complexity of the displacement, squeezing and rotation operators may have interesting applications.
For example, coherent and squeezed states of light are essential building blocks of quantum optics \cite{PhysRevA.31.3093,osti_6703629,Schumaker}, and these states can play an important role in continuous variable quantum computation, see e.g.~\cite{Braunstein:2005zz,Liu_2016,Zhuang:2019jyq}.
For example, in one such algorithm \cite{qumode} the squeezing $r$ is inversely proportional to the precision $\Delta_E$ of phase estimation (for fixed computational time), $r \sim 1/\ln(\Delta_E)$.
Since we found that the complexity is proportional to the squeezing (\ref{eq:CSqueeze2}), this implies that 
the resulting precision for a measurement constructed from a squeezed state with complexity ${\mathcal C}_*$ scales exponentially with the complexity 
$\Delta_E \sim e^{-{\mathcal C}_*}$, suggesting that complexity itself might be an exploitable resource for quantum computation algorithms.
Alternatively, the average energy (or particle number) of a squeezed vacuum state -- another useful quantum information resource -- scales with the squeezing parameter as
\be
\langle E\rangle \sim \bar N = \langle 0|\hat S^\dagger(r,\phi)\hat N \hat S(r,\phi) 0\rangle = \sinh^2 r\, .
\ee
For small squeezing $r \ll 1$, the average energy scales as the square of the squeezing parameter $\langle E\rangle \sim r^2$ so that the complexity scales as the square root of the average energy ${\mathcal C} \sim \sqrt{\langle E\rangle}$.
Interestingly, this dependence of the complexity on the average energy is identical to that of the coherent state (\ref{eq:DispComplexEnergy}).
For large squeezing, however, the average energy scales exponentially with the squeezing $\langle E\rangle \sim e^{2r}$.
The resulting complexity therefore scales logrithmically with the average energy
${\mathcal C} \sim \ln \sqrt{\langle E\rangle}$.
We see here that a benefit to such large-squeezing states is that the quantum circuit complexity needed to build a squeezed state with some average energy $\langle E\rangle_*$ scales slower with $\langle E \rangle_*$ than its coherent state counterpart.
It would be interesting to explore whether this flattening of the dependence of complexity on average energy occurs for other resources, and whether complexity itself can serve as a quantum information resource.

In Section \ref{sec:Applications}, we used our  results for the operator complexity of the quantum harmonic oscillator to study the complexity of a free massive scalar field. In the continuum limit with a UV cutoff $\Lambda$, we found that the complexity rapidly grows linearly with time followed by saturation at a value ${\mathcal C}_\phi \sim L^{d/2} \Lambda^{d/2}$ that depends on the UV cutoff and the number of spatial dimensions.
As an application of our quantum mechanical and field theory results, we studied the operator complexity of quantum cosmological perturbations in a de Sitter background.
The time evolution of Fourier modes of cosmological perturbations can be characterized by two-mode squeezing and rotation operators with time-dependent squeezing and rotation parameters.
Restricted to pairs of Fourier modes, we find that while the operator complexity oscillates rapidly at early times (corresponding to modes deep within the Hubble horizon), at late times the squeezing dominates and the operator complexity grows linearly with the number of e-folds of expansion ${\mathcal C}_{\vec{k}} \sim 2N_e \sim 2\ln a$. Up to the overall numerical coefficient, this agrees with previous results on the complexity of cosmological perturbations at late times obtained using the wavefunction-method complexity and covariance matrix-method complexity \cite{cosmology1,cosmology2}.
Integrating over all Fourier modes, the operator complexity roughly splits into the geometric sum of two regimes: a UV part, which diverges as the free scalar field, and an IR part, which represents growth due to squeezing and grows exponentially as the volume of de Sitter space $\sim H_{dS}^3 e^{N_e}$.
A similar integration for the covariance matrix-method complexity yields a similar IR part but a weaker UV divergence, while the total integrated wavefunction-method complexity is remarkably UV-finite and scales as the (square root of the) volume of de Sitter space.
Together, these results for the total, integrated complexity of quantum cosmological perturbations suggest that the complexity of the observable universe grows by a factor of $e^{3 N_e/2} \sim e^{90}$ during $N_e\sim 60$ e-folds of de Sitter inflation due to the growth in volume.
It would be interesting to consider whether this rapid growth in complexity is consistent with bounds on the growth or saturation of complexity expected from more general considerations (e.g.see \cite{Maldacena2016-mb,Brown:2017jil} for examples), and its relation to the entanglement entropy of cosmological perturbations (see \cite{Brandenberger_1992,Brandenberger_1993,Brahma:2020zpk} for some examples).

Finally, we close with some thoughts on future directions. 
We have focused here the operator geometry generated by simple operators based on the creation and annihilation operators of a quantum harmonic oscillator.
%two simple 3-dimensional Lie groups, the Heisenberg group and SU(1,1), because of their relationship to the algebra of creation and annihilation operators of a single quantum harmonic oscillator.
It would be interesting to extend this analysis to more general systems and their associated groups, including self-interacting and coupled oscillators.
It is also unclear what relationship, if any, there is between the many different approaches towards computing complexity in the literature, even for the simple harmonic oscillator system considered here.
It would also be interesting to consider whether there are physical constraints in which some measures of complexity are better suited to characterizing the ``circuit depth'' cost of construction than others.

\section*{Acknowledgments}

We would like to thank Arpan Bhattacharyya and Sayura Das for helpful conversations and discussions. S.H.~would like to thank the University of Cape Town for funding this project.

%%%%%%%%%%%%%%%%%%%%%%%%%%%%%%%%%%%%%%%%%%%%%%%%%%%%%%%%%%%%%
%%%%%%%%%%%%%%%%%%%%%%%%%%%%%%%%%%%%%%%%%%%%%%%%%%%%%%%%%%%%%
%%%%%%%%%%%%%%%%%%%%%%%%%%%%%%%%%%%%%%%%%%%%%%%%%%%%%%%%%%%%%

% %\begin{thebiVbliography}{99} 

% %\providecommand{\href}[2]{#2}
% \addcontentsline{toc}{section}{References}
% %\bibliographystyle{JHEP}
% \bibliographystyle{utphys}
% \bibliography{nonlocal}   
   
%\end{thebibliography} 

\appendix 

\section{Alternative Matrix Representations}
\label{app:AltRep}

In this appendix, we will choose an alternative matrix representation for the Heisenberg group, establishing representation-independence of the result.

%\subsection{Heisenberg Group}

We begin with the Heisenberg group by noting that the solutions for the operator tangent vectors from
(\ref{eq:HeisenbergEulerArnold}) are independent of the matrix representation of the generators of the Heisenberg group
\be
V^1(s) &=& v_1\, ; \nonumber \\
V^2(s) &=& v_2\, ; \label{eq:AppendHeisEASoln}\\
V^3(s) &=& 0\, , \nonumber
\ee
with a corresponding expression for the complexity (\ref{eq:HeisIntermediateComplexity})
\be
{\mathcal C}_{\rm target} = \sqrt{v_1^2 + v_2^2}\, .
\label{eq:AppHeisIntermediateComplexity}
\ee
We will determine the $v_1, v_2$ constants by matching them to our target operator, which is a solution to
\be
\frac{d\hat U(s)}{ds} = -i V^I(s)\ \hat e_I\ \hat U(s)\, ,
\label{eq:AppHeisUDiffeq}
\ee
subject to the boundary conditions $\hat U(0) = \hat{\mathds{1}}$, $\hat U(1) = \hat {\mathcal U}_{\rm target} = \hat{\mathcal D}(\alpha)$.
In order to find an explicit solution, we need to use a matrix representation of the Heisenberg group.
In the main text, we used a standard 3 x 3 upper-triangular matrix representation of the Heisenberg group generators.
In this appendix, we will instead use a 4 x 4 matrix representation, which takes the form
\be
\hat e_1 = \begin{pmatrix}
	0 & 0 & 0 & 0 \cr
	0 & 0 & 0 & 0 \cr
	0 & -i/2 & 0 & 0 \cr
	-i/2 & 0 & 0 & 0
	\end{pmatrix}\,, \hspace{.2in}
\hat e_2 = \begin{pmatrix}
	0 & i/2 & 0 & 0 \cr
	0 & 0 & 0 & 0 \cr
	0 & 0 & 0 & 0 \cr
	0 & 0 & -i/2 & 0
\end{pmatrix}\,, \hspace{.2in}
\hat e_3 = \begin{pmatrix}
	0 & 0 & 0 & 0 \cr
	0 & 0 & 0 & 0 \cr
	0 & 0 & 0 & 0 \cr
	0 & i/2 & 0 & 0
\end{pmatrix}\,. \label{eq:App4x4HeisRep}
\ee
Within this representation, a general element of this 4 x 4 matrix representation of the Heisenberg group becomes
\be
\hat U(s) = \begin{pmatrix}
	1 & b(s)/2 & 0 & 0 \cr
	0 & 1 & 0 & 0 \cr
	0 & -a/2 & 1 & 0 \cr
	-a/2 & c/2 & -b/2 & 0
\end{pmatrix}\,. \label{eq:App4x4HeisU}
\ee
Using (\ref{eq:App4x4HeisRep}) and (\ref{eq:App4x4HeisU}) in (\ref{eq:AppHeisUDiffeq}) and imposing the boundary conditions, 
the parameterizations $a(s),b(s),c(s)$ of the group generators have the solutions
\be
a(s) &=& v_1 s = -\sqrt{2} {\rm Im}\left[\alpha(t)\right]\, ; \\
b(s) &=& v_2 s = -\sqrt{2} {\rm Re}\left[\alpha(t)\right]\, ; \\
c(s) &=& 0\, .
\ee
The resulting complexity (\ref{eq:AppHeisIntermediateComplexity})
\be
{\mathcal C}_{\rm Heis} = \sqrt{2}\ |\alpha|\, ,
\ee
is identical to the result (\ref{eq:DisplacementComplexity}) obtained using the 3 x 3 matrix representation.
We conclude that the physical result -- the circuit depth -- appears to be independent of the matrix representation, as expected.

\bibliographystyle{utphysmodb}

\bibliography{refs}

\end{document}